\documentclass[amsmath, amssymb, aps, prx, twocolumn, nofootinbib, notitlepage, superscriptaddress]{revtex4-2}
\usepackage[utf8]{inputenc}
\usepackage{graphicx}               % Include figure files
\usepackage[caption=false]{subfig}  % Subfigures
\usepackage{dcolumn}                % Align table columns on decimal point
\usepackage{bm}                     % Bold math symbols
\usepackage[dvipsnames]{xcolor}     % Enable dvips names for color
\usepackage{slashed}                % Feynman slash notation
\usepackage{hyperref}               % Hyperlinked references
\usepackage{enumitem}               % Customizable lists
\usepackage{mathtools}              % Math beautification
\usepackage{floatrow}

\newcounter{myparagraphs}
\begin{document}
\title{Hydrodynamics of the electronic Fermi liquid: a pedagogical overview}
\author{Aaron Hui}
\email{aaron\_hui@brown.edu}
\affiliation{Department of Physics, Ohio State University, Columbus, Ohio 43202, USA}
\affiliation{Department of Physics, Brown University, Providence, RI, 02912, USA}

\author{Brian Skinner}
\affiliation{Department of Physics, Ohio State University, Columbus, Ohio 43202, USA}
\date{\today}

\begin{abstract}
For over a hundred years, electron transport in conductive materials has been primarily described by the Drude model, which assumes that current flow is impeded primarily by momentum-relaxing collisions between electrons and extrinsic objects such as impurities or phonons. In the past decade, however, experiments have increasingly realized ultra-high quality electronic materials that demonstrate a qualitatively distinct method of charge transport called hydrodynamic flow. Hydrodynamic flow occurs when electrons collide much more frequently with each other than with anything else, and in this limit the electric current has long-wavelength collective behavior analogous to that of a classical fluid. 
While electron hydrodynamics has long been postulated theoretically for solid-state systems, the plethora of recent experimental realizations has reinvigorated the field. Here, we review recent theoretical and experimental progress in understanding hydrodynamic electrons using the (hydrodynamic) Fermi liquid as our prototypical example.

\end{abstract}

\maketitle
\section{Introduction}

The past decade has seen a great revival of interest in electron hydrodynamics, following a series of experimental papers in 2016 \cite{Crossno2016, Bandurin2016, Moll2016}. Several reviews have already been written on the subject of electron hydrodynamics \cite{Narozhny2017, Lucas2018review, Narozhny2022, Narozhny2022a, Varnavides2023, Fritz2024}, including one in this journal. Here, we hope to complement these reviews by providing a pedagogical overview of hydrodynamic phenomenology using the Fermi liquid as our prototypical example. As we will discuss, the spirit of hydrodynamics is to provide a universal framework with which to understand transport behavior. 
With this guiding principle, we attempt to balance avoiding material-specific details as much as possible while also avoiding fully general expressions that would obfuscate the presentation. 
Along the way we clarify common points of confusion, including typical abuses of terminology.

After a general presentation of the hydrodynamic approach in the remainder of this section, we review a variety of phenomena associated distinctly with hydrodynamic electrons (Sec.~\ref{sec: fermi liquid hydrodynamics}). Included in Sec.~\ref{sec: fermi liquid hydrodynamics} are a few topics which are not discussed in previous reviews, including the ``paradoxical'' behavior of hydrodynamic electrons in a Corbino geometry, hydrodynamic flow in a smooth potential, and the nature (and utility) of current noise in hydrodynamic electron systems. In Sec.~\ref{sec: beyond} we briefly review phenomena that go beyond the isotropic hydrodynamics of Fermi liquids that is the primary focus of our review, and we conclude in Sec.~\ref{sec: conclusion}.

\subsection{Electron Transport and the Hydrodynamic Regime}
\label{sec: electron transport and the hydro regime}

In electron transport, the Drude model (and its semiclassical extension) has been the model par excellence \cite{Ashcroftbook}. It describes conductivity as the result of non-interacting electrons randomly colliding with obstacles as the electrons are pulled along by an electric field. 
In the classical Drude picture, each collision ``resets'' the electron momentum; this loss of momentum leads to energy dissipation and resistance.
Over distances longer than the mean free path of electrons, the Drude model gives a local Ohm's law
\begin{align}
    \mathbf{E} = \rho_0 \mathbf{J}
    \label{eq: ohm's law}
\end{align}
where $\rho_0 = m\gamma_\text{mr}/(nq^2)$ for $n$ the electron density, $q$ the electron charge, $m$ the electron mass, and $\gamma_\text{mr}$ the scattering rate (more properly called the momentum relaxation rate). Relaxation of the momentum in materials can arise from a number of scattering processes, including electron-impurity scattering, electron-phonon scattering, and electron-electron Umklapp scattering. The momentum relaxation rate is therefore given by $\gamma_\text{mr} \approx \gamma_\text{imp} + \gamma_\text{umklapp} + \gamma_\text{e-ph}$ to first approximation via the Matthiessen's rule. One can show that each of these have different temperature dependencies, namely
\begin{align}
    \rho_0 \sim \rho_\text{imp} T^0 + \rho_\text{umklapp} T^2 + \rho_\text{e-ph} T^{d+2}
    \label{eq: ohmic temperature dependence}
\end{align}
at low temperature, where $d$ is the spatial dimension (for temperatures large compared to the Bloch-Gr\"uneisen temperature, the rate of electron-phonon scattering typically becomes linear in $T$).
Thus, it is generally expected that a conductor has a resistivity that monotonically increases with temperature; in contrast, an insulator's resistivity decreases with temperature as a result of the increase in thermally-activated carriers. A metal is therefore typically defined as a material whose resistivity increases with temperature. 

One can also ask what happens to an ``ideal'' electron system for which the electrons do not undergo any scattering at all. If one makes a pristine channel with negligible phonon and umklapp scattering, the conductance (inverse resistance) in such a system is still finite even though $\gamma_\text{mr} \rightarrow 0$. This no-scattering limit is known as the ballistic transport regime \cite{Datta1995}, and its classical analogue is sometimes called the Knudsen or rarified gas regime. Semiclassically, this regime corresponds to the limit where the mean-free path $\ell_\text{mr}$ for electron momentum relaxation is much greater than the sample size $L$, so that electrons travel unimpeded until they collide with the walls of the sample. In this regime, the \emph{conductance} is given by \cite{Beenakker1991book, Datta1995}
\begin{align}
    G = N\frac{e^2}{2\pi \hbar}
    \label{eq: Landauer conductance}
\end{align}
where $N$ is the number of conducting channels (assuming perfect transmission). The quantization arises due to the finite width of the sample, analogous to the bound states in the infinite square well. Due to the long mean free path, ballistic transport is highly non-local; there is no intrinsic meaning to a conductivity per se, as the conductance is manifestly independent of the sample length. This quantum-mechanical result is due to Landauer\cite{Landauer1957, Datta1995}, with a semiclassical generalization due to Sharvin (called Landauer-Sharvin conductance) \cite{Sharvin1965}.

The hydrodynamic regime sits apart from both the ohmic and ballistic regimes. It was noted as early as 1932 by Peirels \cite{Peierls1932} that not all scattering mechanisms contribute to resistivity, but only those that relax momentum. In particular, electron-electron scattering (without umklapp) preserves the total electron momentum, and therefore it does not relax the current. The hydrodynamic regime is therefore defined by the condition that non-current-relaxing electron-electron scattering dominates over all other scattering mechanisms that do relax the current. More formally, it is defined by $\ell_\text{ee} \ll \ell_\text{mr}, L$ -- the electron-electron scattering length $\ell_{ee}$ is much shorter than all other length scales of interest, including the momentum-relaxing scattering length $\ell_{mr}$ and the sample size $L$. This regime is analogous to the typical situation for classical fluids, such as water or air, whence the name ``hydrodynamics''. 

Approximate momentum conservation in the hydrodynamic regime leads to a host of novel phenomena unusual from the perspective of standard electron transport. For instance, Peierels' argument implies that a perfectly momentum-conserving hydrodynamic regime would have ``infinite conductivity.'' It turns out this argument is a bit too quick; though conductivity is ill-defined, hydrodynamic electrons have a separate dissipative mechanism -- viscosity -- giving rise to a finite conductance in a finite sample with ``rough'' boundaries. Though somewhat foreign in electron transport, these features are mundane in the classical setting. Fluid flow in free space (or a pipe with smooth boundaries) under a uniform force field is clearly dissipationless, while flow through a pipe with rough boundaries exhibits frictional losses. In this review, we detail some of these phenomena and their distinguishability from traditional ohmic and ballistic transport.

\subsection{The Hydrodynamic Framework}
\label{sec: framework}

We take care to distinguish the hydrodynamic regime, as discussed above, from the hydrodynamic framework. The hydrodynamic regime as discussed above amounts to a microscopic specification of the electron-electron scattering rate. By contrast, the hydrodynamic framework is a modeling tool. In the spirit of Landau and Ginzburg, the approach of hydrodynamics is to phenomenologically model a system with a long-wavelength effective theory of (approximately) conserved quantities. This approach is in analogy to the operating principle of thermodynamics, where one uses a handful of macroscopic state variables to summarize and avoid treating an Avogadro's number of microscopic degrees of freedom.As a result, the hydrodynamic framework is used in a wide variety of applications beyond particle and charge transport, including spin dynamics, cold atom systems, and integrable models (sometimes under the name ``generalized hydrodynamics'') \cite{Castro-Alvaredo2016, Bertini2016, Doyon2020}. In fact, hydrodynamics can be thought of as a perturbative dynamical extension of equilibrium thermodynamics, with a similarly universal applicability.

Observables in hydrodynamics are governed by two types of equations: continuity equations and constitutive relations. The continuity equations, as their name suggest, correspond to statements about the local flow of conserved quantities such as charge or momentum. Constitutive relations, on the other hand, are model-dependent relations between observables; equations of state in typical thermodynamics such as the ideal gas law or the linear stress-strain relation (i.e.\ Hooke's law) in linear elasticity theory are classic examples of a constitutive relation.

We can apply hydrodynamic techniques to understand electron transport, just as in classical fluids. In particular, the momentum-conserving hydrodynamic regime is amenable to treatment under the hydrodynamic framework. As a prototypical example, we assume that particle number and momentum are (approximately) conserved. This leads to the following continuity equations.
\begin{align}
    \partial_t n + \partial_i j_i &= 0
    \\
    \partial_t g_i + \partial_j \Pi_{ij} &= - \gamma_\text{mr} g_i .
    \label{eq: momentum continuity}
\end{align}
Here, $n$ and $j_i$ are the particle density and the particle current in direction $i$, while $g_i$ and $\Pi_{ik}$ are the momentum ``density'' (momentum current) and the momentum ``current'' (momentum flux density) \footnote{Here and elsewhere, repeated indices imply a sum; we use Einstein summation notation.}. The term $\gamma_\text{mr} g_i$ in the momentum continuity equation is a phenomenological term commonly added to account for the fact that momentum conservation is only approximate in material systems, equivalent to the Drude resistivity in Eq.~\eqref{eq: ohm's law}. In other words, $\gamma_\text{mr} g_i$ is the rate at which momentum is removed from the electron system in the system's bulk. We assume that momentum relaxation is sufficiently weak so that $g_i$ does not decay to zero on the time scales we care about. 

With $1 + 2d + d^2$ variables (in $d$ dimensions) but only $1 + d$ equations, we need to supplement the continuity equations with model-specific constitutive relations to close the set of equations. 
As a prototypical example of constitutive relations, one could take the following.
\begin{align}
    j_i &\equiv n u_i
    \\
    g_i &\equiv m n u_i
    \\
    \Pi_{ij} &= p\delta_{ij} + qn \phi \delta_{ij} + mn u_i u_j - \sigma_{ik}'
    \label{eq: momentum constitutive relation}
    \\
    \sigma_{ij}' &= \eta \left(\partial_j u_i + \partial_i u_j - \frac{2}{d} \delta_{ij} \partial_k u_k\right) + \zeta \delta_{ij} \partial_k u_k
\end{align}
The first two equations are relatively self-explanatory; the particle current and momentum current are given by number density $n$ or mass density $mn$ (with $m$ the particle mass) multiplied against the hydrodynamic (drift) velocity $u_i$. The momentum flux $\Pi_{ij}$ in Eq.~\eqref{eq: momentum constitutive relation} is a bit more involved; it includes a pressure $p$ and an ``electric pressure'' $qn\phi$, with $q$ the particle charge and $\phi$ the electric potential. Furthermore, there is a convective term $mn u_i u_j$ corresponding to the fact that momentum is carried by the fluid flow; this term arises in the total derivative of $g_i \equiv mnu_i$. Finally, the last term $\sigma'_{ik}$ is the (deviatoric) viscous stress tensor. This tensor describes the effects of viscosity, including the dynamic shear viscosity $\eta$ and dynamic bulk viscosity $\zeta$ in dimension $d$. In the presence of rotational and time-reversal symmetry, viscosity is constrained to these two parameters \cite{landauv6}. We assume for simplicity that the viscosity coefficients are constant. These lead to the Navier-Stokes-Ohm equations supplemented with current continuity.
\begin{align}
    \partial_t u_i + u_k \partial_k u_i & = - \frac{1}{mn}\partial_i p  - \frac{q}{m} \partial_i \phi - \gamma_\text{mr} u_i + \nu \partial^2 u_i
    \nonumber
    \\
    & \phantom{- \partial_i p - } + \left(\tilde{\zeta} + \frac{d-2}{d}\nu\right)\partial_i \partial_k u_k
    \label{eq: Navier-Stokes}
    \\
    \partial_t n + \partial_i (n u_i) &= 0
    \label{eq: current continuity}
\end{align}
where $\nu \equiv \eta/(mn)$ and $\tilde{\zeta} \equiv \zeta/(mn)$ are the kinematic shear and bulk viscosities, respectively. In particular, notice that the Navier-Stokes-Ohm equation [Eq.~\eqref{eq: Navier-Stokes}] is just the momentum equation, or ``$F=ma$,'' while Eq.~\eqref{eq: current continuity} is just the continuity equation for the current. Thus, there are only $d+3$ unknowns in this set of equations: the $d$ components of the hydrodynamic velocity $u_i$, the particle density $n$, the pressure $p$, and the electric potential $\phi$. In particular, we take the mass $m$, the charge $q$, and the viscosities $\nu$ and $\tilde{\zeta}$ to be given as phenomenological parameters. We note that the system of equations is not yet closed; the state variables $p$ and $\phi$ require their own constitutive relations, which are typically set by equilibrium considerations, i.e.\ equations of state. With the exception of the ohmic $-\gamma_\text{mr} u_i$ term, the above equations are structurally identical to those of the hydrodynamics of classical fluids. As we will see, Eq.~\eqref{eq: Navier-Stokes} and Eq.~\eqref{eq: current continuity} form the framework around which much hydrodynamic phenomena are discussed. 

For many situations of interest, the full Eq.~\eqref{eq: Navier-Stokes} and Eq.~\eqref{eq: current continuity} are unwieldy to use. Therefore, we discuss here a number of common simplifications. Oftentimes, one works in the low Reynolds number regime\footnote{The dimensionless (viscous) Reynolds number $\operatorname{Re} = uL/\nu$ characterizes the relative importance of inertial terms in the Navier-Stokes equations. The term $u$ is the typical hydrodynamic velocity and $L$ is the characteristic length scale of the experimental device.}, where the hydrodynamic velocity is relatively small, and drops the nonlinear convective acceleration term $u_k \partial_k u_i$in Eq.~\eqref{eq: Navier-Stokes}. In experimental settings, the Reynolds number for hydrodynamic electrons is of order $10^{-2}$ \cite{Lucas2018review, Narozhny2019, Narozhny2022, Hui2021}. The resulting equation is known as the Stokes-Ohm equation\footnote{``Navier-Stokes'' refers to the nonlinear version.}.

Another common assumption is the incompressible flow (incompressibility) assumption $\partial_i u_i = 0$. This assumption is valid when $u \ll c$, where $c$ is the speed of sound in the fluid. Particularly for the condensed matter audience, it is important to note that the incompressible flow assumption is \emph{different} than stating that the electronic state is incompressible, though the two are related. The latter is an equilibrium statement about the ``stiffness'' $d\mu/dn$ of the electron fluid, where $\mu$ is the chemical potential, which in turn affects the speed of sound. The incompressible flow assumption, on the other hand, is a non-equilibrium statement about the strength of the flow velocity. Therefore, in a strictly incompressible fluid, only incompressible flows are allowed. We also remark that though it is common to think of incompressibility as the assumption $n = \text{const}$, this is not strictly true. The constitutive relations for pressure typically demand a spatial variation in $n$; for a classical incompressible fluid, one keeps the pressure variations while discarding explicit density variations. 

Finally, one often works in steady state, i.e., where all time derivatives are zero. Putting these three assumptions (low Reynolds number, incompressibility, and steady-state) together, the Stokes-Ohm equations of typical interest are
\begin{align}
    - \frac{q}{m} \partial_i \phi  &=  (\gamma_\text{mr} - \nu \partial^2) u_i
    \label{eq: Stokes-Ohm}
    \\
    \partial_i u_i &= 0.
\end{align}
For simplicity, we have dropped the pressure term; under our three assumptions, $p$ behaves identically to $\phi$ and can be innocuously absorbed into $\phi$ (this is not necessarily true of compressible flows \cite{Hui2021}). The great majority of existing experimental realizations of electron hydrodynamics are well-described by this much simplified version of the hydrodynamic equations (we discuss some exceptions in Sec.~\ref{sec: beyond}).

In this form, the Stokes-Ohm equations are very similar to the typical Ohm's law. In fact, one can rewrite Eq.~\eqref{eq: Stokes-Ohm} as
\begin{align}
    - \partial_i \phi = \rho_0(1 - \lambda^2 \partial^2) j_i
    \label{eq: nonlocal ohm's law}
\end{align}
where $\rho_0 \equiv nq^2/(m\gamma_\text{mr})$ is the Drude resistivity and $\lambda \equiv \sqrt{\nu/\gamma_\text{mr}}$ is referred to as the Gurzhi length. Physically, the Gurzhi length has the meaning of the length scale over which one can think of the flow profile as hydrodynamic. The kinematic shear viscosity $\nu$ has the units of a diffusion constant, and indeed one can think of $\nu$ as the microscopic diffusion constant of a single particle as it undergoes collisions with other particles. The particle loses its momentum to extrinsic scattering (with impurities or phonons) on a time scale $\tau_\text{mr} \sim \gamma_\text{mr}^{-1}$, and in this time scale it diffuses a distance $\lambda \sim \sqrt{\nu \tau_\text{mr}}$. The Gurzhi length $\lambda$ should therefore be thought of as the distance over which particle flow is momentum-conserving; over distances much longer than $\lambda$ the flow is ohmic.

In Fourier space, it becomes clear that viscosity is nothing but the $k^2$ component of a non-local conductivity $\rho(\mathbf{k})$, and therefore the Stokes-Ohm hydrodynamics is an expression of a non-local Ohm's law. As a result, macroscopic transport features describable by Eq.~\eqref{eq: nonlocal ohm's law} do not necessarily have a hydrodynamic origin. This distinction has led to some debate over what should constitute hydrodynamic behavior, with some papers instead referring to non-local transport behavior described by Eq.~\eqref{eq: nonlocal ohm's law} without an electron-electron scattering origin as a form of ``modified hydrodynamics.'' \cite{Hui2020, Aharon-Steinberg2022, Wolf2023, Estrada-Álvarez2024} 

Conversely, one can be in a hydrodynamic regime yet exhibit mundane behaviors. For instance, viscosity could be negligibly small or one could work in a very large sample of size $\ell$ such that $\lambda \ll \ell$. This inequality is not necessarily inconsistent with the assumption that $\gamma_\text{ee} \ll \gamma_\text{mr}$ as is necessary for the hydrodynamic regime. In such a case, the flow profiles are fully described by the usual Ohm's law except within a small ``boundary layer'' region of size $\sim \lambda$ near the sample boundaries.
%; any unique signatures of the hydrodynamic regime would not be captured by the hydrodynamic approach. 
To emphasize situations where $\lambda$ is non-trivially large, such situations are sometimes referred to as ``viscous hydrodynamics.''

As a closing remark, much confusion arises from the conflation of the hydrodynamic regime, a microscopic statement about $\gamma_\text{ee}$, and the hydrodynamic framework, a long-wavelength effective description. Although the hydrodynamic framework is powerful in its universality and its ability to treat strongly-correlated systems, it relies on phenomenological parameters that must be supplied externally. As such, microscopic features relevant to the hydrodynamic regime, e.g.\ the functional form of viscosity, cannot be treated by the hydrodynamic framework. Throughout this paper, we emphasize the distinction between these two whenever possible, reserving the term electron hydrodynamics to refer to the hydrodynamic regime. Furthermore, we will refrain from describing the equations of motion as the ``hydrodynamic equations'' to avoid potential confusion, as they need not arise from strong electron-electron scattering.

\subsection{Why Hydrodynamics Now?}

With the concept of hydrodynamics dating back at least to the 1800s, the concept of a hydrodynamic electron fluid is certainly not new. Hydrodynamics of the electron gas was first studied by Bloch \cite{Bloch1934} and Jensen \cite{Jensen1937} in the 1930s, only some 30 years after the discovery of the electron by J.J. Thomson \cite{Thomson1897}. They emphasized the utility of hydrodynamics to understand the oscillatory ``bulk collective response'' modes of the interacting electron system, where the fluid formally lived in an infinite, unbounded region. From their work, hydrodynamics was invoked to study, e.g., the various sound modes in liquid He-3 or plasmonic modes in metals \cite{Tomonaga1950, Ritchie1957, Nozieres1958, Bennett1970, Ying1974, Quinn1976, DasSarma1979, DasSarma1982, Pines2018book1, Nozieres2018book2}.

The idea of hydrodynamic electrons in metals as a ``confined'' fluid, of primary interest for this review, is often first associated with Gurzhi's work in the 1960s, who emphasized finite-size and boundary effects \cite{Gurzhi1963, Gurzhi1968}. Since then, these ideas have cycled in and out of favor. With the experimental realization of 2D electron gases in the late 1970s, interest in hydrodynamic flow was reignited. Of particular note is Dyakonov and Shur's work in 1993 \cite{Dyakonov1993}, where they proposed that the nonlinearities present in the ideal hydrodynamic equations of motion (Euler equations) could lead to the so-called Dyakonov-Shur instability to generate THz waves. In 1995, de Jong and Molenkamp \cite{deJong1995} observed hydrodynamic effects in (Al,Ga)As wires. Most recently, a flurry of activity has emerged in the past decade following the discovery of hydrodynamic behavior in a number of ultraclean materials \cite{Crossno2016, Bandurin2016, Moll2016}. Monolayer graphene has been the primary material of interest \cite{Crossno2016, Bandurin2016, Ghahari2016, KrishnaKumar2017, Sulpizio2019, Ku2020, Berdyugin2019, Gallagher2019, Kim2020, Kumar2022, Jenkins2022, Samaddar2021, Krebs2023}, though hydrodynamic signatures have also been reported in other 2DEG systems \cite{deJong1995, Gusev2018, Levin2018, Braem2018, Keser2021, Gupta2021, Gusev2021, Vijayakrishnan2023, Palm2024}, Weyl semimetals \cite{Gooth2018, Jaoui2018, Vool2021}, the delafossite PdCoO$_2$ \cite{Moll2016, Bachmann2022, Baker2024}, and Sb \cite{Jaoui2021}.

The seasonality of interest in electron hydrodynamics is most certainly linked to experimental accessibility. In large part, until recently it has proven exceedingly difficult to find a material in which the electron system can realize the hydrodynamic regime. At a minimum, the mean free path $\ell_\text{mr}$ for momentum relaxation must be larger than the electron-electron scattering length $\ell_{ee}$. Even in what one would consider good metals, such as copper or gold, one finds $\ell_\text{mr} \sim 40$ nm at room temperature \cite{Gall2016}. By contrast, current experiments for the best graphene samples also estimate $\ell_\text{ee} \sim 40$ nm \cite{Lucas2018review}. As a result, only recently have materials been discovered that are sufficiently clean and free of electron-phonon scattering to exhibit hydrodynamic behavior, and even in these situations one can often achieve the condition that $\ell_\text{ee}$ is moderately smaller than $\ell_\text{mr}$. As such, usually one must keep momentum-relaxation effects in order to accurately model experiments.

\section{Fermi Liquid Hydrodynamics}
\label{sec: fermi liquid hydrodynamics}

The prototypical example of electron hydrodynamics is exemplified by a Fermi liquid. As an (effective) weakly-interacting system with well-defined quasiparticles, one can use Boltzmann kinetic theory to explicitly derive the hydrodynamic equations of motion and the constitutive relations. Fermi liquid theory, though often applied to diffusive (disordered) metals, was originally developed to describe liquid He-3, from which parameters such as viscosity and thermal conductivity were computed and tested against experiment \cite{Abrikosov1959}. We emphasize that it is important to distinguish the ``clean'' Fermi liquid of liquid He-3 from the ``metallic'' Fermi liquid typically discussed in materials. While both are amenable to a Fermi liquid theory treatment, the metallic Fermi liquid degrades momentum rapidly from, e.g.\ disorder, phonon, or umklapp scattering and does not lie in the hydrodynamic regime. By contrast, liquid He-3 and the various experimental materials mentioned in the previous section have relatively weak momentum relaxation such that hydrodynamic transport phenomena can be observed.

For simplicity, we consider a quadratic quasiparticle dispersion $\epsilon_k = \hbar^2 k^2/(2m)$, where $m$ is the quasiparticle mass. To compute observables, one utilizes the Boltzmann kinetic equation
\begin{align}
    \left[ \partial_t + \mathbf{u} \cdot \nabla_\mathbf{r} + \left( q\mathbf{E} + \frac{q}{c} \mathbf{u}\times\mathbf{B}\right) \cdot \nabla_\mathbf{k}  \right] f = I[f]
    \label{eq: Boltzmann}
\end{align}
where $f(\mathbf{r},\mathbf{k},t)$ is the semiclassical quasiparticle distribution function and $I[f]$ is the collision integral. The form of the collision integral depends on the various scattering processes the quasiparticles undergo. Since $I[f]$ is generically difficult to compute, one often makes simplifying assumptions. One common assumption is
\begin{align}
    I[f] = I_{ee}[f] + \frac{f_0 - f}{\tau_\text{mr}}
    \label{eq: collision integral assumption}
\end{align}
where $I_{ee}$ is the momentum-conserving electron-electron interaction, while the second term is a relaxation-time approximation of momentum-relaxing collisions with $f_0$ the equilibrium distribution. There is an implicit assumption made here that the two scattering mechanisms can be separated like this; one assumes that the two scattering mechanisms can be treated independently in a Matthiessen-rule way. It is important to remark that this assumption, while convenient, is not necessarily true. This assumption implies that non-local viscous transport must arise from electron-electron interactions, but as discussed in Sec.~\ref{sec: framework} this is not necessarily the case. We continue to comment on this point throughout Sec.~\ref{sec: fermi liquid hydrodynamics}.

The Boltzmann formalism allows us to derive the continuity relations discussed above. Collisional invariants, i.e.\ quantities conserved under collisions, allow us to circumvent explicit calculation of the collision integrals. These invariants $\mathcal{O}(\mathbf{k})$ are defined as quantities such that
\begin{align}
    \int d\mathbf{k} \,  \mathcal{O}(\mathbf{k}) \, I[f](\mathbf{k}) = 0
\end{align}
i.e., they vanish upon integration against the collision integral. Integrating the Boltzmann equation [Eq.~\eqref{eq: Boltzmann}] multiplied by some collisional invariant $\mathcal{O}$ over momentum $\mathbf{k}$ removes the pesky collision integral. In the absence of fields ($\mathbf{E} = \mathbf{B} = 0$), what remains is a continuity equation for the object $\int dk\, \mathcal{O} f \equiv n \langle O \rangle_f$, i.e.\ the density of $\mathcal{O}$ given the distribution $f$. For example, setting $\mathcal{O} = 1$ corresponds to $\int d\mathbf{k}\, \mathcal{O} f = n$ the particle density. Thus, integrating both sides of the kinetic equation gives
\begin{align}
    \partial_t n + \nabla_\mathbf{r} \cdot (n\mathbf{u}) = 0,
\end{align}
which reproduces the density continuity equation with $j_i = nu_i$. With more work, one can also show that the contributions from finite $\mathbf{E}$ and $\mathbf{B}$ terms also vanish (see e.g.\ \cite{Narozhny2019}). Similarly, by setting $\mathcal{O} = k_i$ one can rederive the momentum continuity equation for each direction $i$. In this case, $k_i$ is a collisional invariant for $I_\text{ee}$ specifically, but one must continue to treat the momentum relaxation $\tau_\text{mr} = 1/\gamma_\text{mr}$ explicitly, resulting in the right-hand side of Eq.~\eqref{eq: momentum continuity}. 

The constitutive relations can also be derived from the Boltzmann approach. As these require computing observables, doing such derivations requires an explicit solution for the distribution function $f$. While a general solution of the Boltzmann equation is difficult, one can tackle it perturbatively, e.g., using the Chapman-Enskog method \cite{Landauv10, Chapman1990book, Cercignani1987book}. This perturbative expansion is controlled by the Knudsen number $\ell_\text{mfp}/L \ll 1$, where $\ell_\text{mfp}$ is the characteristic mean free path (including both momentum-conserving and momentum-relaxing collisions) and $L$ is the characteristic sample size.

As a technical remark, in this review we will only concern ourselves with the frequent collision regime $\omega \tau \ll 1$, where $\omega$ is the frequency of the applied field and $\tau \sim v_F \ell_\text{mfp}$ is a characteristic scattering time \footnote{We remark that since we are working in a finite sample, the limits $\omega \tau \ll 1$ and $\ell_\text{mfp}/L \ll 1$ are different. In the study of bulk collective modes, often $L\rightarrow \infty$ so $\omega\tau$ is the only relevant scale of interest. Somewhat confusingly, both $\omega \tau \gg 1$ and $\ell_\text{mfp}/L \gg 1$ are called ``collisionless'' regimes. However, we emphasize that these are physically distinct; in particular, $\ell_\text{mfp}/L \gg 1$ requires a careful treatment of boundary scattering, in contrast to $\omega \tau \gg 1$ which can be treated as bulk physics. For perturbative expansions about $\omega\tau \gg 1$ (and discussions regarding the $\omega\tau$ crossover from first to zero zound), see e.g. Ref.~\cite{Pines2018book1, baym1991book, Chapman1990book, Landauv10, Cercignani1987book, Magner2014, Magner2017}.}. Except when explicitly stated, we take $\omega \rightarrow 0$ and work in steady-state. At present, the cleanest samples have $1/\tau_\text{mr} \sim 0.1$ to $1$ THz \cite{Bandurin2016, Hui2021}, making it difficult to reach $\omega \tau \gg 1$. Furthermore, most experimentally observed phenomena operate with DC sources. As such, this low-frequency choice is often the experimentally relevant one. 

To simplify the discussion, we temporarily ignore momentum relaxation and set $\tau_\text{mr} \rightarrow \infty$. The idea of the derivation is as follows. As a zeroth-order ansatz, we guess a distribution function $f_0(\mathbf{r},\mathbf{k}, t)$ such that the collision integral $I[f_0] = 0$. At equilibrium where $\mathbf{v} = \mathbf{E} = \mathbf{B} = 0$, it is clear that $f_0$ is an exact solution. This ansatz is equivalent to assuming local equilibrium; at each point, local collisions alone cannot alter the distribution function. This can be visualized by dividing the fluid into subsystem parcels at the coarse-graining scale, where each subsystem is assumed to establish thermodynamic equilibrium. With $f_0$ in hand, one can make explicit calculations of macroscopic observables via integration against $f_0$, e.g., $\int dk \mathcal{O} f_0$. Therefore, one can use the microscopic definitions of various observables via $f_0$ to explicitly derive the constitutive relations; this is analogous to computing equations of state in equilibrium thermodynamics. If one stops at the zeroth-order ansatz $f_0$, any relaxational dynamics of collisions are completely ignored since $I[f_0] = 0$. This results in the Euler equations, i.e.\ hydrodynamics without dissipation.

To obtain dissipative coefficients, we need to keep track of how perturbations $\delta f = f - f_0$ decay back to local equilibrium. Formally, such calculations require us to perturbatively solve the (linearized) Boltzmann equation for $\delta f$, which involves explicitly treating $I[f]$ as well as the left-hand side (the ``streaming'' terms) of Eq.~\eqref{eq: Boltzmann}. In doing so, one finds corrections to observables due to integration against $f = f_0 + \delta f$ and therefore to the constitutive relations. For example, computation of the momentum-flux tensor $\Pi_{ij}$ using $f$ leads to an additional correction, namely the the viscous stress tensor $\sigma'_{ij}$ (see Eq.~\eqref{eq: momentum constitutive relation}). In particular, one finds that $\delta f$ gives the first-order terms responsible for dissipation, e.g.\ the shear and bulk viscosities. Turning momentum-relaxation back on adds the ohmic dissipation term to the momentum equation. See, e.g., Ref.~\cite{Landauv10, Chapman1990book, Cercignani1987book} for more details on the Chapman-Enskog approach or Ref.~\cite{Narozhny2019} for a modern application to graphene.

As we see, the Boltzmann approach allows us to rigorously derive the hydrodynamic equations for sufficiently weakly-interacting systems and provides an ab-initio approach to computing the constitutive relations and phenomenological parameters used in the hydrodynamic approach. Furthermore, as a microscopic approach, it allows us to study other transport regimes in a single framework, such as the ballistic and ohmic transport regimes and their crossovers into the hydrodynamic regime. One can also prescribe microscopic descriptions of what happens at sample boundaries rather than impose them phenomenologically, and in so doing ``derive'' the boundary conditions. This is done by modifying the collision integral appropriately to reflect the regime of interest. \cite{Briskot2015, Narozhny2019, Kiselev2019, Raichev2022, Lucas2017a, Lucas2018a, Lucas2018b, Shytov2018, Cook2019, Valentinis2023, Levchenko2020, Sulpizio2019, Holder2019a}.

In the remainder of this section, we discuss various hydrodynamic phenomena associated with the clean Fermi liquid. One of the main challenges of identifying transport regimes experimentally is the inability to directly measure the rate of microscopic scattering mechanisms. Consequently, studies of electron systems rely on indirect methods to infer hydrodynamic flow, and identifying uniquely hydrodynamic signatures is not always obvious. Many phenomena first attributed to hydrodynamic behavior were later found to be realizable in other transport regimes. Therefore, we will note the distinctiveness (or lack thereof) of each phenomenon, comparing it with the usual ohmic and/or ballistic regimes when appropriate. 

\subsection{Flow Profiles}
\label{sec: flow profiles}

We begin by studying the linear flow regime using the Stokes-Ohm equation [Eq.~\eqref{eq: Stokes-Ohm}]
\begin{align}
    - \partial_i \phi = \rho_0(1 - \lambda^2 \partial^2) j_i,
    \label{eq: Stokes-Ohm flow profile}
\end{align}
reproduced here for convenience \footnote{A classical analogue for the Stokes-Ohm equation is fluid flow through a porous media. Flow in this media experiences drag, which could be modeled by $\gamma_\text{mr}$.}. The $\lambda \rightarrow 0$ limit is identical to Ohm's law, while the Gurzhi length $\lambda$ controls deviations from ohmic flow. The regime $\lambda \gtrsim 1$ is sometimes called ``viscous hydrodynamics,'' where viscous corrections to transport become significant. The presence of $\lambda$ turns the algebraic Ohm's law into the differential Stokes-Ohm equation, significantly increasing the complexity of the model. Flow patterns become scale-dependent, and a proper solution to the flow profile requires knowledge of the boundary conditions on $j_i$. 

\begin{figure}
    \centering
    \includegraphics[width=\linewidth]{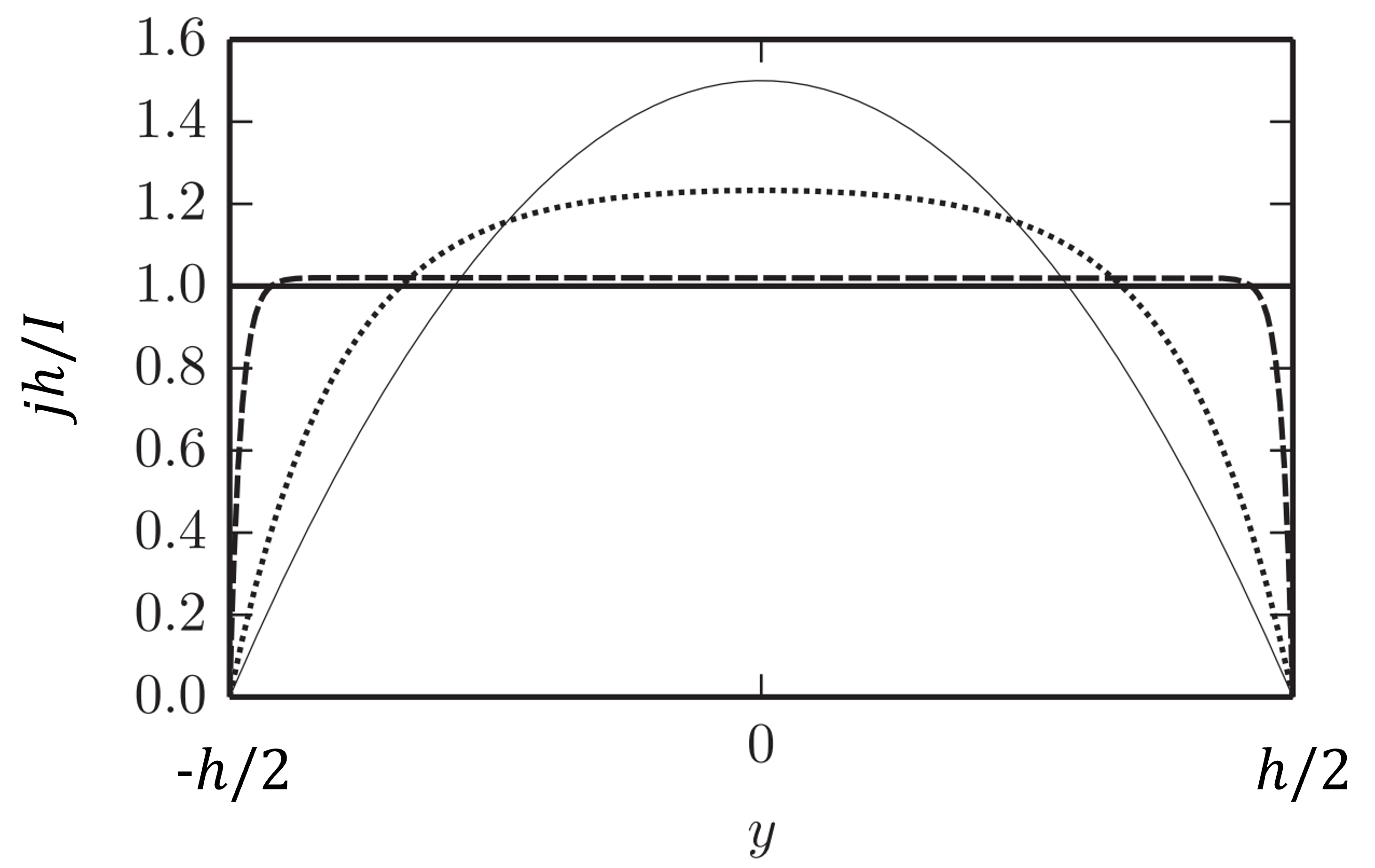}
    \caption{A plot of the normalized current density $j$ that arises from a solution to the Stokes-Ohm equation in a rectangular channel with no-slip boundary conditions. $h$ is the width of the channel and $I$ is the total current. The thin line corresponds to $\lambda/h \rightarrow \infty$, the dotted line to $\lambda/h = 0.1$, the dashed line $\lambda/h = 0.01$, and the thick solid line (uniform flow) $\lambda/h = 0$. Adapted from Ref.~\cite{Torre2015}.}
    \label{fig: stokes-ohm flow}
\end{figure}

Consider first the classic case of hydrodynamic flow in the simple setting of a rectangular channel with no-slip boundary conditions, see Fig.~\ref{fig: stokes-ohm flow}. In this case, the Stokes-Ohm equation is exactly solvable \cite{Torre2015}. Tuning $\lambda$ from 0 to $\infty$ interpolates between the rectangular ``ohmic'' profile and the parabolic Poiseuille profile. The Gurzhi length $\lambda$ describes the length below which viscous effects are important; in this case, it is approximately the distance from the boundary over which the flow profile has appreciable curvature. However, the shape of the profile is sensitive to the choice of boundary conditions. In the limiting case of a stress-free boundary condition, the profile is always spatially uniform regardless of the value of $\lambda$. Without a good understanding of the boundary physics, it is difficult to draw sharp conclusions from flow profile data.

Other geometries can further amplify non-trivial curvature in flow. Judicious choices can lead to eddies and whirlpools at sufficiently large viscosity \cite{Levitov2016, Pellegrino2016, Falkovich2017, Aharon-Steinberg2022, Palm2024}. These topological features cannot appear in ohmic flow since the electric potential satisfies a Laplace equation; harmonic functions can only take their minimum on the boundary of the domain, but a vortex exhibits a stagnation point in the sample bulk. In ohmic flow, the current is always proportional to the electric field and therefore cannot exhibit a stagnation point. By contrast, in viscous flow the current need not be proportional to the electric field. As an experimental signature, the observation of whirlpools has the advantage of being qualitatively distinct.

Unusual flow curvatures, including the observation of Poiseuille-like flow and vortices, have been experimentally observed. They have been measured through direct imaging techniques \cite{Sulpizio2019, Ku2020, Jenkins2022, Aharon-Steinberg2022, Palm2024} as well as non-local resistance measurements \cite{Bandurin2016, Bandurin2018}. However, one must be cautious in attributing these observations to hydrodynamic effects \cite{Pellegrino2016}. As discussed earlier, the Stokes-Ohm equation is completely equivalent to a non-local resistivity. A non-local resistivity (e.g. ``viscosity'') need not arise from strong electron-electron interactions. Several papers have pointed out the possibility of non-local flow behavior in purely disordered \cite{Hui2020} and ballistic regimes \cite{Shytov2018, Chandra2019, Nazaryan2024}, far away from a true hydrodynamic regime. Furthermore, a low viscosity does not imply that electron-electron interactions are not strong. In the clean Fermi liquid, for instance, a higher rate of electron-electron collisions (e.g.\ at higher temperature) implies a smaller viscosity, since a faster rate of collisions reduces the diffusion constant. Furthermore, for large samples of $L \gg \lambda$ or in situations when flow gradients are small, viscous flow effects might not be macroscopically observable. Thus, a hydrodynamic material may display pedestrian ohmic transport, as mentioned above. In general, multiple tests of the hydrodynamic regime are needed, which is a running theme of this section.

\subsection{Electrical Resistance}

Electrical resistance in a hydrodynamic material is controlled by two dissipative coefficients: the momentum relaxation rate $\gamma_\text{mr}$ and the (kinematic shear) viscosity $\nu$.\footnote{For time-dependent flows, one also needs to consider the bulk viscosity $\zeta$. In general, viscosity is a rank-4 tensor [see Eq.~\eqref{eq: full viscosity}].} In the linear flow regime, we saw in Sec.~\ref{sec: framework} that the Stokes-Ohm equation [Eq.~\eqref{eq: Stokes-Ohm}] is equivalent to Ohm's law with a non-local resistivity $\rho(\mathbf{k})$. While viscous dissipation is fundamentally nonlocal in nature, one can think that in regions of space where the current profile has a spatial gradient, there are two parallel sources of resistivity: one coming from momentum relaxation and another from viscous shear. For example, if we approximate the second derivative on the right-hand side of Eq.~\eqref{eq: Stokes-Ohm flow profile} by a negative constant $\sim -1/L^2$, where $L$ is the typical length scale over which the current varies, then we see that the effect of viscosity is essentially to add a second parallel channel for dissipation in a Matthiessen-rule way: $\gamma_\text{tot} \sim \gamma_\text{mr} + \nu/L^2 = \gamma_\text{mr} (1 + \lambda^2/L^2)$. Physically, one can think that viscous shear adds a second parallel mechanism by which heat is dissipated by the current flow.

As we saw in Sec.~\ref{sec: flow profiles}, for a sample with finite momentum relaxation rate $\gamma_\text{mr}$, viscous effects are restricted to a ``viscous boundary layer'' of some thickness $L$. If we want to estimate the resistance $R$ of a channel with width $h$ and length $\ell$, we can first estimate the dissipated power from the current flow. This dissipated power has two sources: a momentum relaxation process that occurs uniformly throughout the sample with power density $j^2 \rho_0$, where $j = I/h$ is the current density and $\rho_0$ is the Drude resistivity, and a viscous dissipation with power density $j^2 \rho_0 \lambda^2/L^2$ that occurs only in the boundary layer. Thus the total power $P$ is given by the sum of the two dissipation sources multiplied by their corresponding area, $P \sim (j^2 \rho_0) \cdot h \ell + (j^2 \rho_0 \lambda^2/L^2) \cdot L \ell$. Equating this sum with $I^2 R$ gives
\begin{align}
    R \sim \frac{\ell}{h} \frac{m [\gamma_\text{mr} + \nu/(Lh)]}{nq^2},
    \label{eq: resistance estimate}
\end{align}
where $m$ and $q$ are the quasiparticle mass and charge, and $n$ is the particle density. Thus both $\gamma_\text{mr}$ and $\nu$ contribute to the total resistance. 
We emphasize that, when dealing with hydrodynamic behavior, it is more accurate to discuss resistance rather than resistivity due to the nonlocality of viscous dissipation. 

A striking consequence of our estimate in Eq.~\eqref{eq: resistance estimate}, which exemplifies the notion that a local resistivity is not well defined in the viscous regime, is that $R$ has qualitatively different scaling with the channel width in the viscous regime as compared to the ohmic regime. Specifically, in the viscous regime where $\gamma_\text{mr} \ll \nu/h^2$ (equivalent to $\lambda \gg h$), the boundary layer becomes as wide as the channel itself, so that $L \sim h$ and consequently $R \sim h^{-3}$. This result is in contrast to ohmic transport, where $R \sim h^{-1}$. The difference in scaling can be understood as follows. In ohmic transport, the electron's momentum is lost as it collides with an impurity, which occurs uniformly at every point in the sample and is width-independent. In viscous hydrodynamic transport, the electron system only loses momentum when electrons collide with the rough boundary; electron-electron collisions only diffuse momentum, and the kinematic viscosity is precisely the diffusion coefficient for momentum. Individual electrons therefore perform a random walk, and a diffusing electron near the middle of the channel on average takes a time of order $h^2/\nu$ to hit the wall and lose its momentum. Thus, the constant momentum relaxation time in the Drude resistivity is replaced by an effective momentum relaxation time that is proportional to $h^2$, so that the resistance acquires an additional factor $1/h^2$.

Experimentally, anomalous non-ohmic width scaling has been observed in various experiments \cite{Moll2016, Gooth2018}, though we caution that as an effect that depends only on the non-locality of resistivity it need not arise from hydrodynamic electron behavior. These cautionary remarks were particularly pertinent to the hydrodynamic candidate PdCoO$_2$, where experiments had observed anomalous width scaling as evidence of hydrodynamic behavior \cite{Moll2016}, but upon further inspection concluded that the experiment was likely not in the hydrodynamic regime \cite{Cook2019, Nandi2018, Baker2024}.

\subsection{Viscosity}

One can ask about the functional dependencies of the dissipative parameters $\gamma_\text{mr}$ and $\nu$. These can be computed from microscopics using, e.g., the Boltzmann formalism. The behavior of $\gamma_\text{mr}$ has long been a quantity of interest in electron transport; details were discussed above around Eq.~\eqref{eq: ohmic temperature dependence}. The viscosity $\nu$ has also long been studied for the clean Fermi liquid, though not in the context of electron transport. By dimensional analysis, kinematic viscosity can be estimated by
\begin{align}
    \nu \sim u\ell_\text{mfp}
    \label{eq: viscosity estimate}
\end{align}
where $u$ is a characteristic velocity and $\ell_\text{mfp}$ is a characteristic scattering length (mean free path) \cite{Landauv10}. We also remark on the somewhat counterintuitive fact that viscosity diverges as $\ell_\text{mfp} \rightarrow \infty$ (while keeping the hydrodynamic assumption $\ell_\text{mfp} \ll L$), implying that weakly-interacting gases have very high viscosity. Under the lens that kinematic viscosity is a diffusion constant, this divergence can be rephrased as stating that particles in a weakly-interacting gas travel a long distance before colliding with another particle and diffusing their momentum. In other words, weakly-interacting particles in a medium with no other scattering mechanisms are able to quickly convey their momentum toward boundaries, so that viscous dissipation is more effective \footnote{We assume here the size of the medium $L \gg \ell_\text{ee}$ so that the hydrodynamic framework is still valid and viscosity has an unambiguous meaning. See the subsequent discussion on apparent or effective viscosity in this section.}.

\begin{figure}
    \centering
    \includegraphics[width=0.9\linewidth]{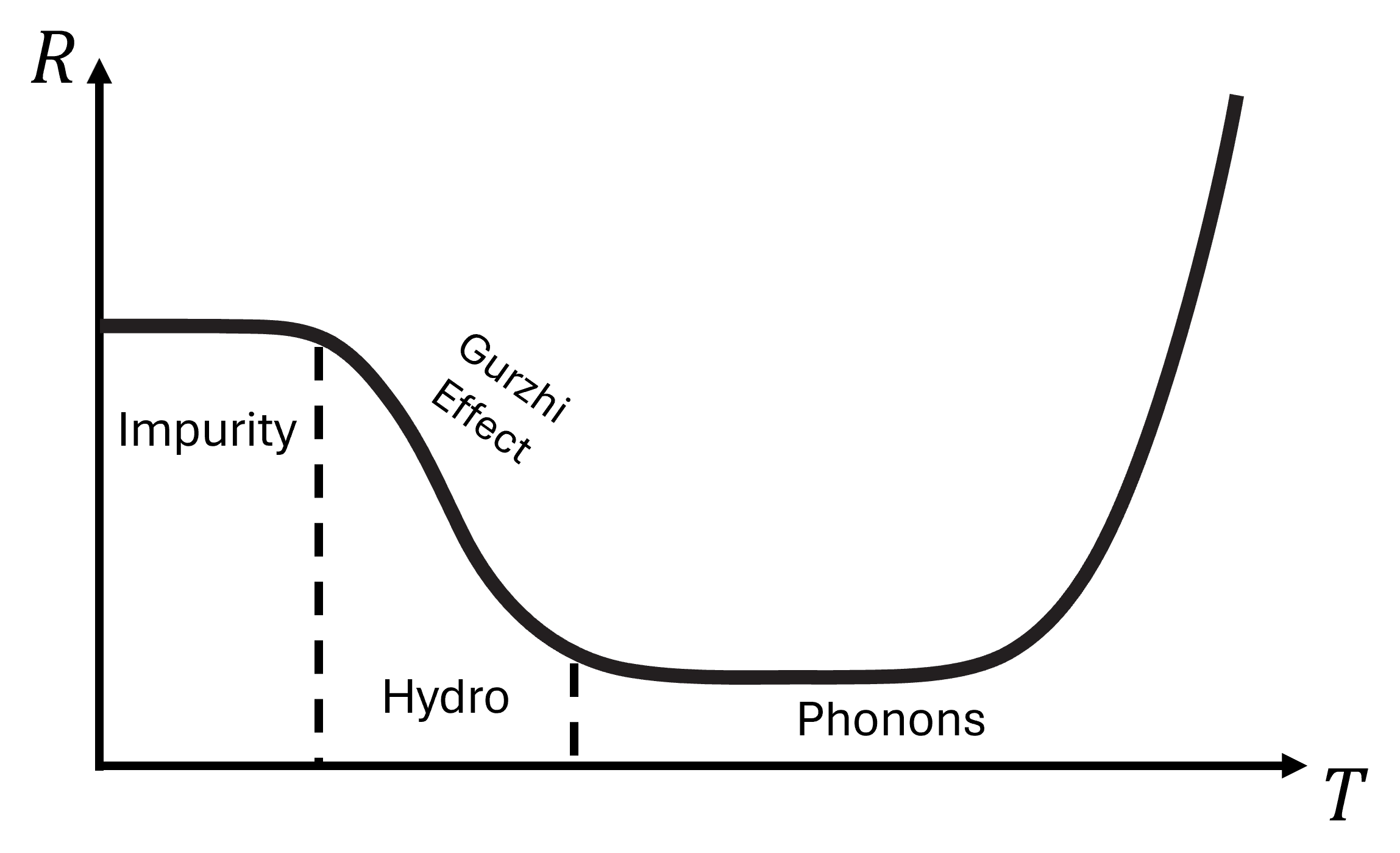}
    \caption{A cartoon of the Gurzhi effect. At low temperatures, impurity scattering dominates and the resistance is temperature-independent. As the temperature increases, eventually $\gamma_\text{ee} \gg \gamma_\text{imp}$ and the material becomes hydrodynamic. Since the viscosity decreases with temperature, so too does the resistance. This decrease is known as the Gurzhi effect. As the temperature increases past the Bloch-Gr\"unesein temperature, electron-phonon scattering becomes dominant, the material leaves the hydrodynamic window, and resistance increases again. Diagram courtesy of Felicia Setiono.}
    \label{fig: Gurzhi Effect}
\end{figure}

For the 3D clean Fermi liquid, we estimate $\ell_\text{mfp} \sim \ell_\text{ee} \sim v_F \tau_\text{ee} \sim v_F E_F/T^2$, so that the kinematic viscosity
\begin{align}
    \nu \propto v_F^2 \frac{E_F}{T^2}
\end{align}
where $E_F = \hbar^2 k_F^2/(2m)$ is the Fermi energy, $v_F =\hbar k_F/m$ is the Fermi velocity, and $T$ is the temperature. Here and throughout, we are using units such that $k_B = 1$. This estimate for $\nu$ can be explicitly verified by a Boltzmann calculation for a clean 3D Fermi liquid \cite{Abrikosov1959, Brooker1968, HogardJensen1968, Sykes1970}, and has been experimentally validated in liquid He-3 \cite{Huang2012}. The electron-electron scattering rate $\gamma_\text{ee} = \tau_\text{ee}^{-1} \sim T^2/E_F$ arises from 2-to-2 scattering around a broadened Fermi surface \cite{Ashcroftbook}. We emphasize that this dependence of $\nu$ on $T$ implies $R \propto 1/T^2$ in the regime of viscous hydrodynamics, and in particular that electrical resistance decreases with temperature in a clean Fermi liquid. This temperature dependence was first studied in depth by Gurzhi \cite{Gurzhi1963, Gurzhi1968}, who considered a weakly-disordered Fermi liquid, arguing that there is a non-trivial temperature window in which the resistance should decrease with temperature. As shown in Fig.~\ref{fig: Gurzhi Effect}, this window is bounded from below by the dominance of impurity scattering and bounded from above by the dominance of electron-phonon scattering. This nonmonotonic temperature dependence defies the traditional categorization of metals and insulators in terms of their temperature-dependence resistivity. Indeed, a clean Fermi liquid, i.e.\ the limit of a clean and perfect metal without phonon scattering, exhibits ``insulator-like'' behavior.

The above estimates also hold for the clean 2D Fermi liquid, albeit with some caveats. As is usual for 2D physics, one finds logarithmic corrections associated with small-angle scattering \cite{Novikov2006, Alekseev2020}. Perhaps more dramatically, recent Boltzmann calculations \cite{Laikhtman1992, Ledwith2019, Ledwith2019b, Alekseev2020, Hofmann2022, Hofmann2023, Hong2024, Rostami2024} have shown that kinematic constraints lead to a distinction in the decay rates between even and odd mode distortions of the distribution function. The system exhibits so-called ``tomographic dynamics'', in which the viscosity becomes scale-dependent. At long wavelengths, one recovers the typical $\nu \propto 1/T^2$ Fermi liquid behavior, but at shorter wavelengths one finds $\nu \sim 1/T$ \cite{Kryhin2024}. As a result, resistance also decreases with temperature in 2D, although with some quantitative differences in the scaling behavior.

While it is conceptually convenient to think of viscosity as arising from momentum-conserving electron-electron interactions and momentum relaxation as arising from impurity or phonon scattering, this description is not quite true. Even in the absence of electron-electron interactions, one can generate a viscous force in the equations of motion\footnote{We use the force term as the operational definition of viscosity. One can alternatively define the viscosity via a stress-strain response, or equivalently by a stress-stress correlation function. These definitions agree in the clean limit, but may differ when momentum relaxation is introduced \cite{Hui2020, Burmistrov2019, Bradlyn2012}.}; since the Stokes-Ohm equation [Eq.~\eqref{eq: Stokes-Ohm}] is equivalent to a non-local Ohm's law, one only needs a mechanism to generate a non-local resistivity in order to produce a viscous force. As a result, using the estimate of Eq.~\eqref{eq: viscosity estimate}, it is clear that all scattering lengths contribute to $\ell_\text{mfp}$; even point-like impurities generate a non-trivial $k^2$ component of resistivity \cite{Alekseev2016, Hui2020, Shytov2018, Chandra2019, Nazaryan2024}. One can approximate $\ell_\text{mfp}^{-1} \sim \sum_i \ell_i^{-1}$ in a Matthessian rule way, demonstrating that the shortest length scale controls the viscosity. This rule is also known in the classical fluids literature, where the apparent or effective viscosity is cut off by the sample width $h$ as one moves towards the ballistic (Knudsen) regime; boundary scattering (with a rate $\propto h^{-1}$), dominates over interparticle scattering (with a rate $\propto \ell_\text{ee}^{-1}$) in the limit of very weak interactions \cite{AliBeskok1999, Hadjiconstantinou2006, Lilley2008, Michalis2010, Zhang2012}. Both theory \cite{Mihajlovic2009, Shytov2018, Hui2020} and experiment \cite{Bandurin2018, Sulpizio2019, Aharon-Steinberg2022} have demonstrated that non-hydrodynamic samples can still develop viscous transport behavior. Conversely, interactions can modify the expected transport behavior induced by momentum-relaxation as opposed to a na\"ive Matthessian rule picture. Most dramatically, interactions can ``lubricate'' electron flows around obstacles and enhance sample conductances \cite{Guo2017, KrishnaKumar2017, Li2020, Li2022, Krebs2023}, see Sec.~\ref{sec: superballistic}. 

\begin{figure}
    \centering
    \includegraphics[width=0.95\linewidth]{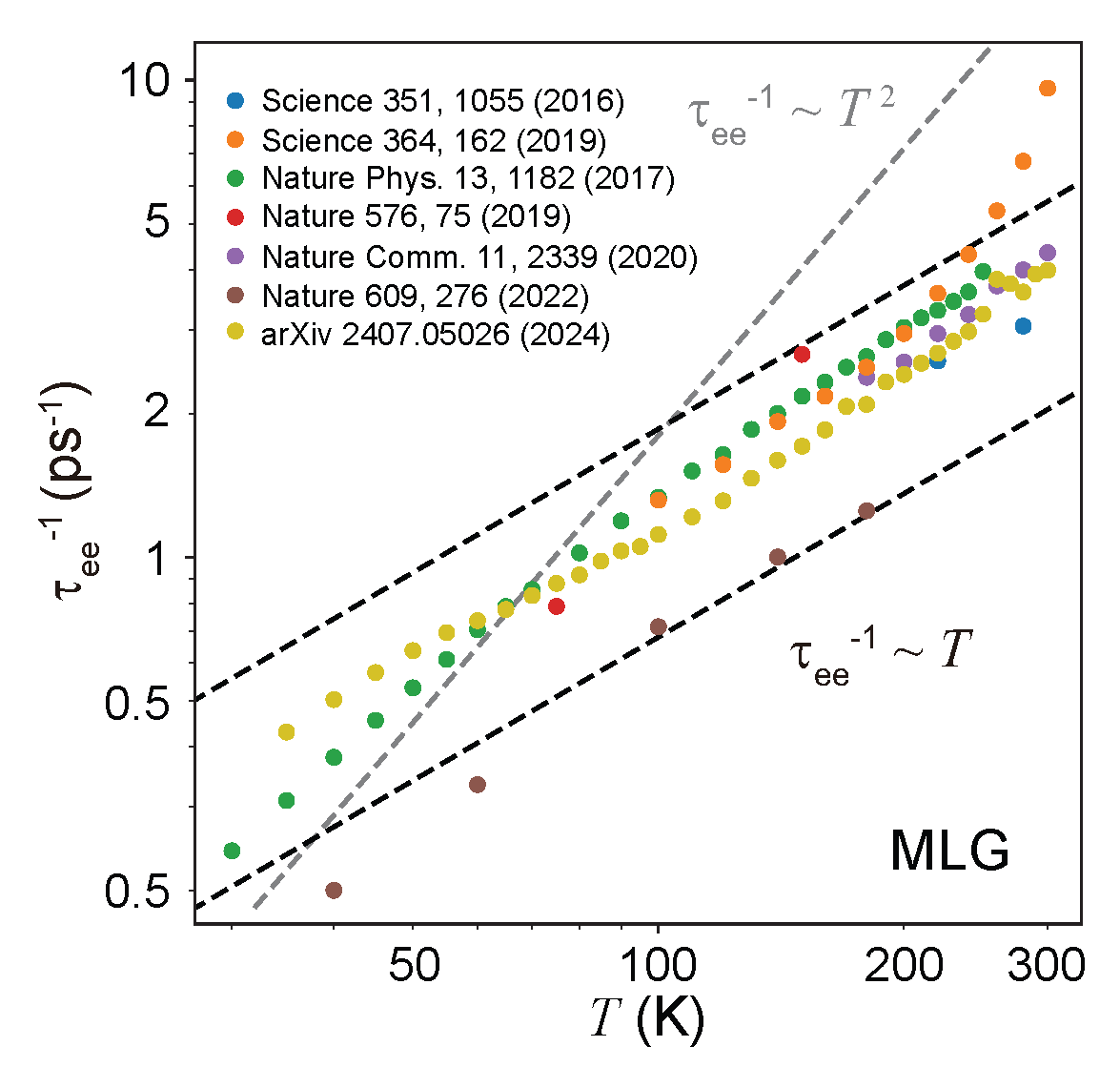}
    \caption{A plot of electron-electron scattering rate $\tau_\text{ee}^{-1}$ (or inverse viscosity $\nu^{-1} \propto \tau_\text{ee}^{-1}$) against temperature for monolayer graphene across a number of experiments. Adapted from a figure courtesy of Yihang Zeng and Cory Dean.}
    \label{fig: mlg viscosity vs temp}
\end{figure}

Unfortunately, experimental evidence validating Fermi liquid predictions for viscosity and the Gurzhi effect remain sparse. Measurements of electron transport in hydrodynamic devices show an increase of resistance with temperature \cite{Bandurin2016, Crossno2016, Zeng2024, Berdyugin2019, Keser2021, Gooth2018, Jaoui2018, Jaoui2021, Thuillier2025}, with the exception of \cite{Gusev2021, Vijayakrishnan2023, deJong1995} and point-contact geometries \cite{KrishnaKumar2017, Kim2020}. Furthermore, viscosity measurements are non-trivial due to the need to disentangle momentum-relaxation contributions. Measured viscosities generally decrease with temperature \cite{Bandurin2016, KrishnaKumar2017, Levin2018, Berdyugin2019, Kim2020, Zeng2024}, but the exact functional form is still unknown; depending on the experiment, the temperature dependence can range from $\nu \propto T^{-2}$ to $\nu \propto T^{-1}$ (see Fig.~\ref{fig: mlg viscosity vs temp}). The story becomes murkier if one looks at the density dependence: some experiments show that viscosity decreases with density \cite{Bandurin2016, Sulpizio2019, Ku2020, Talanov2024}, others show an increase with viscosity with density \cite{KrishnaKumar2017, Kim2020, Zeng2024}, and some show no dependence on density \cite{Zeng2024}. In general, transport measurements of viscosity are difficult due to the need to isolate the viscous component of transport, and only a handful of such measurements have been done. Much remains to be done, both experimentally and theoretically, for a complete picture of viscosity to emerge.

\subsection{Superballistic transport}
\label{sec: superballistic}

One dramatic signature of non-trivial hydrodynamic behavior is so-called ``superballistic transport''. As discussed in Sec.~\ref{sec: electron transport and the hydro regime}, the conductance in the ballistic regime is given by the Landauer formula [Eq.~\eqref{eq: Landauer conductance}]
\begin{align}
    G = N \frac{e^2}{2\pi \hbar}
\end{align}
where $N$ is the number of open ballistic channels in the sample and $h$ is Planck's constant. In electron transport lore, this ballistic conductance acts as a ``fundamental'' upper bound; even in a sample with no impurities or phonons (a ``perfect'' metal), resistance will be quantum-limited by the width of the device. If the width of the constriction $w$ is larger than the Fermi wavelength $\lambda_F = 2\pi/ k_F$, then one can approximate $N$ as $w/\lambda_F$, i.e.\ the number of ``non-overlapping'' electrons that can fit through the constriction. This approximation leads to the semiclassical Landauer-Sharvin formula \cite{Sharvin1965, Guo2017}
\begin{align}
    G = \frac{ge^2}{2\pi\hbar} \frac{w}{\lambda_F}
\end{align}
where $g$ is the degeneracy of the momentum states (spin, valley, etc.). 

\begin{figure}
    \centering
    \includegraphics[width=.7\linewidth]{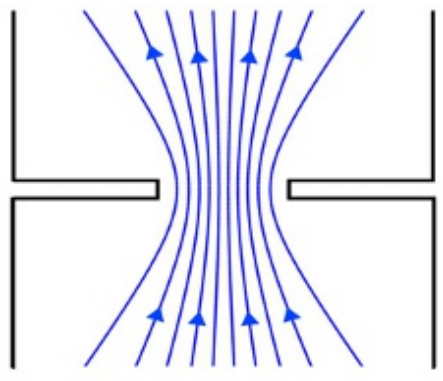}
    \caption{A sketch of viscous flow through a constriction geometry. Viscosity couples adjacent layers of flow, pulling them through the constriction. This effect enables a superballistic conductivity. Adapted from Ref.~\cite{KrishnaKumar2017}.}
    \label{fig: superballistic}
\end{figure}

However, the above results rely on the absence of interactions. In the hydrodynamic regime, one can get around this ``fundamental'' upper bound; interactions can reduce resistance \cite{Guo2017, Stern2022, Nagaev2008, Nagaev2010, Melnikov2012}. This effect was demonstrated most prominently in a constriction or ``point-contact'' geometry. For ballistic transport, a simple classical picture is that electrons only travel in straight lines. Therefore, only electrons that are aligned with the constriction can pass through, while all others are reflected. For hydrodynamic transport, on the other hand, adjacent layers of fluid are coupled. Fluid passing through the constriction can drag adjacent layers of fluid through the constriction through a viscous force, enabling previously ballistically-forbidden fluid layers to pass through, see Fig.~\ref{fig: superballistic}. The viscous coupling enhances the conductance of the constriction geometry, allowing for conductances higher than what is allowed by the Landauer-Sharvin formula. This effect is known as superballistic transport. It was experimentally observed in Refs.~\onlinecite{KrishnaKumar2017, Kim2020, Kumar2022, Krebs2023, Kravtsov2024, Ginzburg2021, Estrada-Alvarez2024}; in point-contact geometries, the conductance was observed to exceed the Landauer-Sharvin expectation by more than $15\%$. The observation of superballistic transport serves as a strong signature of interaction-dominated flow.

\subsection{The Corbino Geometry and the Corbino Paradox}
\label{sec: corbino}

\begin{figure}
    \centering
    \includegraphics[width=.9\linewidth]{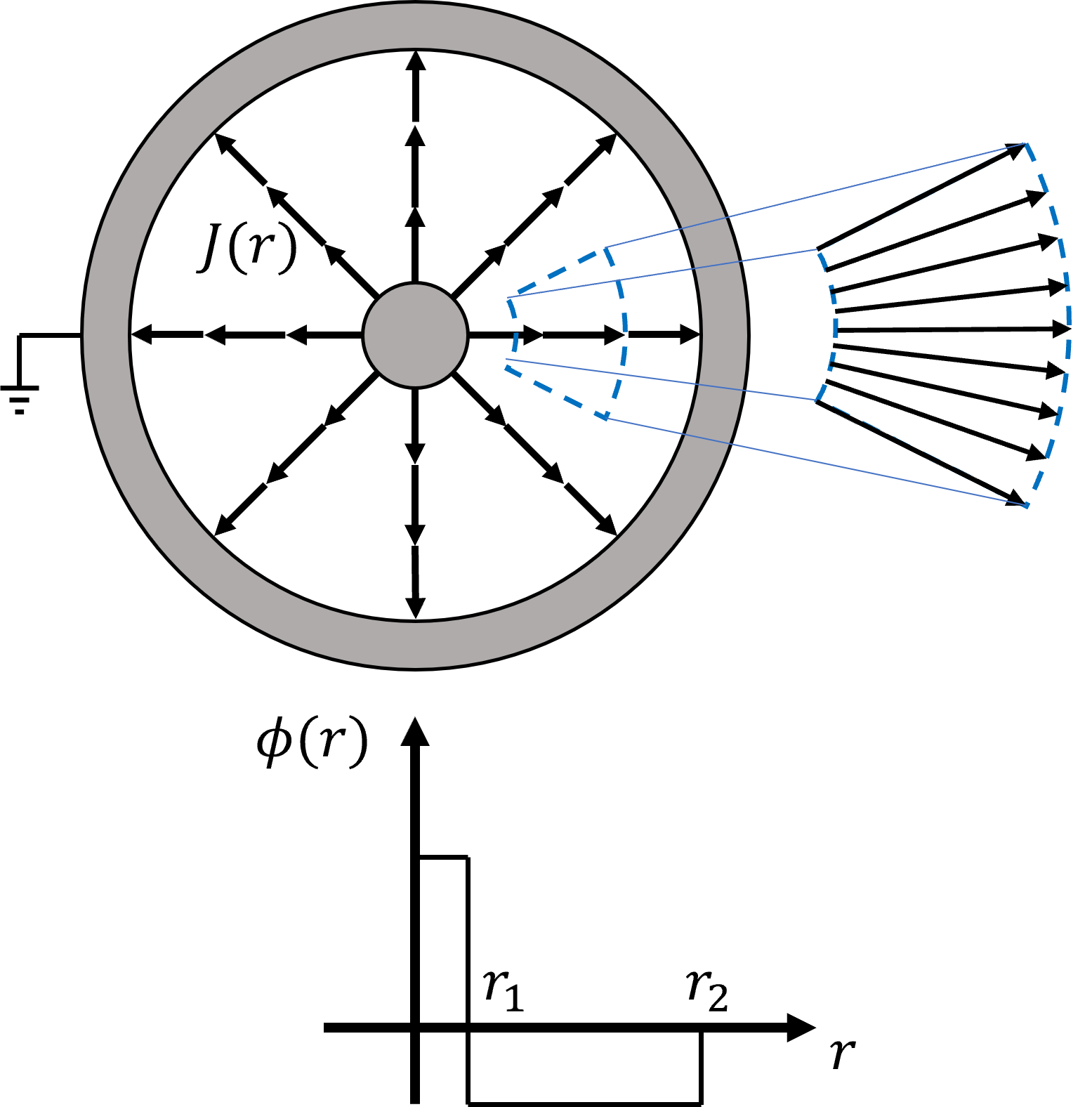}
    \caption{A plot of current flow and electric potential in the Corbino geometry in the hydrodynamic limit. The expanding flow implies a non-zero viscous stress and therefore non-trivial viscous heating. Despite this production of heat, the electric potential is constant in the bulk of the device (i.e., there is no driving electric field). This apparent inconsistency is known as the Corbino paradox. It is resolved by the sharp drops in voltage at the inner and outer contacts. Adapted from Ref.~\cite{Hui2022} and Ref.~\cite{Shavit2019}.}
    \label{fig: corbino}
\end{figure}

Due to the geometry-dependent nature of many hydrodynamic effects, it is useful to explore different device configurations. One particularly interesting geometry is the Corbino or annular geometry, depicted in Fig.~\ref{fig: corbino}. In the absence of momentum relaxation, attempting to solve the Stokes equation [Eq.~\eqref{eq: Stokes-Ohm}] immediately leads to subtleties. Rotational symmetry enforces that $u_r \propto I/r$, where $I$ is the total current. However, $1/r$ is a harmonic function and therefore the viscous force term $\nu \nabla^2 \mathbf{u}$ vanishes. Therefore, there is no voltage drop throughout the electron system even in the presence of a finite current. The situation becomes even more strange when one notices that although the viscous force vanishes, the viscous heating rate is non-zero. This leads to a paradoxical state of affairs, first raised by Ref.~\cite{Shavit2019} -- there is no local $\mathbf{E} \cdot \mathbf{J}$ power input anywhere, but everywhere there is energy loss through dissipative heating. We call this seeming inconsistency the ``Corbino paradox'' \cite{Hui2022}.

This paradoxical behavior is a reflection of the non-locality of viscous hydrodynamic behavior. In ohmic transport, both the $\mathbf{E}\cdot\mathbf{J}$ energy input and the dissipation of energy occur at the same point in space. However, in viscous hydrodynamics the energy input only occurs at the boundary where the electrons are pushed into the sample. Once inside the Corbino disk, the electrons dissipate their energy through viscous shearing. The energy loss must be reflected in the power dissipated by the battery; there must be a non-zero voltage drop from the source to the drain contact. To reconcile the lack of a voltage drop across the hydrodynamic material and the non-zero voltage drop across the battery, there must be a sharp contact voltage drop between the metallic leads and the hydrodynamic sample \cite{Shavit2019, Stern2022, Raichev2022, Hui2024}, see Fig.~\ref{fig: corbino}.

Let us reformulate the above argument, but from a constructionist perspective. When one discusses a voltage across a sample, they specifically have in mind the voltage drop as measured at the metallic leads. However, in a hydrodynamic material, the interfacial physics becomes critical because of the presence of non-locality. This is analogous to ballistic Landauer-Sharvin physics, where the finite energy dissipation occurs at the interface between the sample and the contact. With the assumption of continuity of energy flux density across the sample-contact boundary, the contact voltage drop is \cite{Shavit2019, Hui2024}
\begin{align}
    \phi^c - \phi = -\frac{2m\nu}{q} K(\hat{\mathbf{n}})(\mathbf{u}\cdot\hat{\mathbf{n}})
\end{align}
where $\phi^c$ is the contact voltage, $\phi$ is the sample voltage, $m$ is the mass, $\nu$ is the kinematic viscosity, $q$ is the charge, and $K(\hat{\mathbf{n}})$ is the signed extrinsic curvature (inverse radius of the osculating circle) relative to the normal $\hat{\mathbf{n}}$. Qualitatively, one can envision that the viscous stress-energy stored in the boundary curvature within the hydrodynamic sample is converted into a finite voltage in the metallic lead \cite{Hui2024}. A key lesson of the Corbino paradox is that one must be much more careful in treating hydrodynamic problems with curved boundaries \cite{Hui2023, Hui2024, Talanov2024}. The Corbino paradox predictions of a contact voltage drop and the vanishing bulk electric field have been experimentally observed, though much care was taken to isolate only the viscous hydrodynamic component to demonstrate these effects \cite{Kumar2022}.

\subsection{Thermal and thermoelectric transport}
Thus far, we have focused solely on electronic transport. In addition to flows of charge and momentum, one may also be interested in the transport of energy. Within the hydrodynamic and Boltzmann frameworks, energy transport is treated in much the same way as charge and momentum. One formulates a continuity equation for energy density, supplementing them with constitutive relations. The continuity equation reads
\begin{align}
    \partial_t n_E + \nabla \cdot \mathbf{j}_E = 0
    \label{eq: energy continuity}
\end{align}
where $n_E$ is the energy density and $j_E$ is the energy current (energy flux density). The constitutive relations for the clean Fermi liquid can again be derived from the Boltzmann kinetic equation \cite{Chapman1990book, Landauv10, Cercignani1987book}. For simplicity, we take the constitutive relations
\begin{align}
    n_E &= nm \left(\frac{1}{2} u^2 + \epsilon\right) \\
    j_{E,i} &= nm \left(\frac{1}{2} u^2 + \epsilon + \frac{P}{nm}\right) u_i - \sigma'_{ij} u_j + \gamma_\text{mr} u^2 +  Q_i
    \label{eq: jE} \\
    \mathbf{Q} &= -\kappa \mathbf{\nabla} T
\end{align}
The energy density $n_E$ is comprised of kinetic energy and the internal energy, with $\epsilon$ the internal energy per unit mass. 
There are four additional terms in $\mathbf{j}_E$. One can think of the energy current as arising from particles carrying energy into and out of a hypothetical closed surface; in this picture, the pressure term $P$ corresponds to work done on the surface due to pressure forces. The third term in Eq.~\eqref{eq: jE} corresponds to a contribution from dissipative viscous processes; the transfer of momentum via viscous processes also results in a transfer of energy. The fourth term is the ohmic dissipation term. The last term $\mathbf{Q}$ is the energy current due to thermal conduction alone, and is typically what we associate with the heat flux in the heat equation governing thermal conduction with thermal conductivity $\kappa$. For simplicity, we assume that $\kappa$ is a scalar constant, although of course in general it can be a rank-two tensor. 

It is often convenient to rewrite the energy continuity equation. By using the momentum continuity equation and standard thermodyamic relations, one can rewrite Eq.~\eqref{eq: energy continuity} into an entropy continuity or heat continuity equation \cite{landauv6}; rather than keeping track of the total energy, one only keeps track of the heat flows corresponding to irreversible dissipative losses which increase the entropy. For an incompressible fluid, this process yields
\begin{align}
    \rho c_p \left(\partial_t T + u_i \partial_i T\right) = \kappa \partial^2 T + \sigma'_{ij} \partial_i u_j + \rho \gamma_\text{mr} u^2
    \label{eq: heat continuity}
\end{align}
Here, the LHS corresponds to the continuity of heat density $\rho c_P T$, where $\rho = m n$ is the mass density and $c_p$ is the specific heat at constant pressure, while terms on the RHS correspond to the sources of heat generation. In the absence of flow $u \equiv 0$, this equation reduces to the heat equation. 

With the thermal transport equation [Eq.~\eqref{eq: heat continuity}] in hand, one can study the interplay between hydrodynamic flow and heat transport in materials. A number of works have studied thermoelectric effects in hydrodynamic systems using the Boltzmann formalism; the same techniques used to calculate viscosity can be used for thermal conductivity and the thermoelectric tensor. In general, hydrodynamic flow entails significant deviations from the Mott relation between thermopower and conductivity and the Wiedemann-Franz law between thermal and electrical conductivity \cite{Muller2008, Foster2009, Lucas2016, Lucas2018, Ghahari2016, Li2020, Zarenia2019, Zarenia2019b, Zarenia2020, Li2022, Pongsangangan2022} (see Sec.~\ref{sec: WF law}), which are hallmarks of the usual metallic Fermi liquid \cite{Ashcroftbook}.

The main hydrodynamic modification to thermal transport [Eq.~\eqref{eq: heat continuity}] is the addition of a viscous heating channel. The new heating channel depends on flow gradients, and therefore depends sensitively on boundary conditions and sample geometry. The local heating power is no longer equal to $\mathbf{E} \cdot \mathbf{J}$ as in the ohmic case; this additional source of heating can dramatically modify the heating profile under current bias, and can in turn non-trivially influence electronic transport \cite{Hui2023, Hui2024, Talanov2024, Andreev2011, HuiPozderac2024, Tikhonov2019, Li2020, Zhang2021, Furtmaier2015}. These effects include nonlinear convective effects such as the Rayleigh-B\'enard instability \cite{Furtmaier2015} and dissipation asymmetries \cite{Tikhonov2019, Zhang2021}. However, experimental works on hydrodynamic thermal transport remain relatively sparse, in part due to the difficulty of conducting thermal transport experiments. In Sec.~\ref{sec: noise} below we discuss how random fluctuations of the electrical current can be used to infer thermal transport details of hydrodynamic electrons.

\subsection{Wiedemann-Franz Law}
\label{sec: WF law}

For a metallic Fermi liquid, both the electrical $\sigma$ and thermal $\kappa$ conductivities \footnote{The thermal conductivity here corresponds only to the electron contribution; phononic contributions are neglected.} are controlled by a single time scale: the momentum relaxation time $\tau_\text{mr}$ (see Eq.~\eqref{eq: collision integral assumption} with $I_{ee} \equiv 0$). As a result, the ratio
\begin{align}
    L_\text{WF} \equiv \frac{\kappa}{\sigma T} \cong \frac{\pi^2}{3} \frac{k_B^2}{e^2}
    \label{eq: WF}
\end{align}
is a universal constant at temperatures $T$ much smaller than the Fermi energy, where $e$ is the fundamental electric charge; $L_\text{WF}$ is known as the Lorenz number. Eq.~\eqref{eq: WF} is known as the Wiedemann-Franz (WF) law, and is derived in the relaxation-time approximation \cite{Ashcroftbook}. Significant departures from the WF law, therefore, signify non-(metallic)-Fermi liquid physics and can serve as supplementary evidence for hydrodynamic behavior.

Before we discuss the WF law in the context of hydrodynamics (i.e. turning on $I_{ee} \neq 0$), however, we need to deal with some subtleties. As previously discussed, the local conductivity $\sigma$ is an ill-defined quantity in hydrodynamics. There are therefore two ways to define the WF ratio. The first way is to ignore viscosity entirely and only take the ohmic contribution to conductivity, namely $\sigma = nq^2/(m\gamma_\text{mr})$. This treatment is equivalent to computing the WF ratio for uniform flow in an infinite domain, and it is a common definition in theory papers. The second way is to define the WF ratio with macroscopic conductances, namely $L_\text{WF} \equiv G_\text{th}/(G_\text{el} T)$, where $G_\text{th}$ and $G_\text{el}$ are the thermal and electrical conductances, respectively. This second definition includes viscous corrections to $G_\text{el}$ and corresponds to the typical experimental approach. While these two approaches agree in the ohmic regime, they are distinct in viscous hydrodynamics; $G_\text{el}$ depends on both the viscosity and the geometry of the sample. This dependence is is exemplified for the clean Fermi liquid with $\gamma_\text{mr} = 0$; in a finite geometry with no-slip boundary conditions, the first definition diverges while the second remains finite.

Regardless of the choice of definition, hydrodynamic systems violate the WF law. This violation has been computed theoretically \cite{Muller2008, Principi2015, Lucas2018, Zarenia2019, Ahn2022} and observed experimentally \cite{Crossno2016, Jaoui2018, Gooth2018, Jaoui2021, Talanov2024} with up to order-of-magnitude violations. For a clean Fermi liquid, the WF ratio is expected to be suppressed. The WF violation can be qualitatively understood as follows. Due to momentum conservation, electron-electron collisions do not degrade the flow of electron current, allowing the electrical conductance to be large. However, electron-electron collisions do not conserve the energy current (though they do conserve total energy) and thus they reduce the thermal conductivity. This decoupling of the electrical and thermal transport times can lead to a significant violation of the WF law in hydrodynamic materials, and in particular a suppression for a clean Fermi liquid. This is sometimes handwavingly presented as $G_\text{th} \sim (\gamma_\text{mr} + \gamma_\text{ee})^{-1}$, but $G_\text{el} \sim \gamma_\text{mr}^{-1}$; we emphasize that this approximation, though helpful, should not be taken too literally due to various subtleties such as those described above.
(As we discuss below in Sec.~\ref{sec: dirac}, the opposite is true in the ``Dirac fluid'' regime, where electrons and holes are present in nearly equal number: electron-hole collisions degrade the electric current but not the energy current, so that $L_\text{WF}$ is substantially enhanced from the WF value.)

\subsection{Hydrodynamic flow in a smooth potential landscape}

Thus far, we have only discussed two sources of momentum relaxation. The first is boundary scattering off of rigid walls. These are objects with hard boundaries that are large compared to the mean free path $\ell_\text{ee}$ for electron-electron collisions, and which give rise to frictional (e.g.\ no-slip) boundary conditions and effects like the Stokes paradox \cite{Hruska2002, Guo2017a, Lucas2017a, Kiselev2019, Raichev2022}. The second is momentum-relaxing scattering, e.g.\ from disorder. In this sense we have considered disorder only as a contributor to the finite momentum-relaxation rate $\gamma_\text{mr}$. Such a description is valid when the disorder potential that produces electron scattering varies over length scales that are much shorter than $\ell_{ee}$. 

In this section, we discuss a third situation: flow in a smoothly varying potential. In this case, the current density adjusts itself continuously in response to the potential, and one cannot reduce the effects of the potential simply to a uniform momentum relaxation rate. Similarly, current flows ``over'' regions of varying potential, and the potential cannot be reduced to a set of ``hard-wall'' obstacles. Due to the slowly-varying nature of the potential, on the hydrodynamic coarse-graining scale $\ell_{ee}$ the potential is approximately constant; we remark that even a strong potential can be treated within the kinetic theory approach so long as the variation is slow with respect to $\ell_{ee}$. Such smooth disorder potentials arise naturally in 2D electron systems due to stray charges embedded in the substrate or in a nearby delta-doping layer \cite{Ando1982, Skinner2013}. Moir\'{e} potentials in twisted or lattice-mismatched 2D systems also provide a smooth potential, which is periodic rather than random \cite{andrei2021marvels}. We comment below on the distinction between periodic and random potentials in terms of their effect on hydrodynamic electrons.

In Ref.~\onlinecite{Andreev2011}, the authors studied the resistivity associated with hydrodynamic electrons flowing through a smooth potential, and they pointed out that the energy dissipation associated with resistivity arises from two competing effects. First, viscous dissipation is proportional to the squared spatial gradient of the fluid velocity (i.e.\ $\propto \eta (\partial_i u_j)^2$), so that viscous forces favor a spatially uniform flow profile. Second, disorder causes the equilibrium electron density to vary in space, which implies variations in the electronic entropy density. Current flowing across an entropy gradient produces an opposing voltage via the thermoelectric effect, which leads to energy dissipation. Thus, thermoelectric forces favor current to be strongly concentrated along equipotential contours, where the entropy density is constant. The optimal flow pattern can be thought of as that which minimizes the total power dissipated (or, more precisely, which minimizes the entropy generated within the sample \cite{HuiPozderac2024}).

When the potential is weak, the current is nearly uniform and the resistivity corresponds to an average of both the thermoelectric dissipation and the viscous dissipation, which are both proportional to the mean-square disorder amplitude \cite{Andreev2011}. However, when the potential is strong, the current concentrates into narrow channels that follow percolating potential energy contours. The optimal width $h$ of these channels arises as a balance between viscous and thermoelectric forces, with the former favoring large $h$ and the latter favoring small $h$. For periodic potentials (but not random potentials), this optimization gives an ``effective'' resistivity \cite{Andreev2011, HuiPozderac2024}
\begin{equation}
\rho_\text{eff} \sim \frac{1}{e^2} \sqrt{ \frac{T \eta m^2 (\delta s)^2}{\kappa n^2 \xi} },
\label{eq: rhoperiodic}
\end{equation}
where $\delta s$ is the typical spatial fluctuation in entropy density (at equilibrium) and $\xi$ is the spatial period of the potential. Ref.~\cite{HuiPozderac2024} calculated the numeric prefactors of this equation for different types of periodic potentials and pointed out that, if one assumes Fermi liquid scaling of the viscosity ($\eta \propto T^{-2}$), thermal conductivity ($\kappa \propto T^{-1}$), and entropy density ($\delta s \propto T$) \footnote{We remark that the notation for $\delta s$ differs from that used in e.g. Ref.~\cite{Landauv9}. Here, $\delta s$ refers to the typical spatial variation of the entropy density as a result of the statically-imposed external potential, while in Ref.~\cite{Landauv9} it refers to fluctuations associated with oscillatory modes. As a result, the relation $\delta s \propto T$ is just the statement of how the entropy density of a Fermi liquid in static equilibrium changes with temperature.}, then Eq.~\eqref{eq: rhoperiodic} implies a linear-in-$T$ behavior of the resistivity. This hydrodynamic flow mechanism is therefore a candidate explanation for the linear-in-$T$ resistivity observed in various strongly-correlated 2D moir\'{e} materials \cite{polshyn2019large, Cao2020, ghiotto2021quantum}. 

\begin{figure}[htb]
    \centering
    \includegraphics[width=0.9\linewidth]{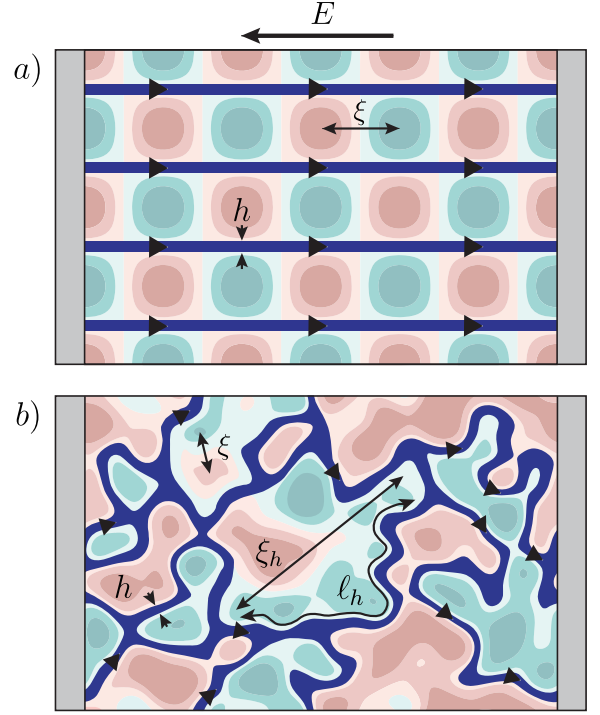}
    \caption{A device sketch with the imposed external potential $U$ (blue and red contour lines), where the shaded dark blue lines correspond to electron flow under the applied electric field $E$. The thin current channels of width $h$ are concentrated about equipotential contours of $U$. a) The case of square periodic potential. b) The case of a random potential, for which the equipotential contours and the current channels meander and become tortuous in nature. Reproduced with permission from Ref.~\cite{HuiPozderac2024}.}
    \label{fig: Flow in strong potential}
\end{figure}

For random potentials, the behavior of the resistivity is complicated by the fact that current-carrying paths become increasingly tortuous as they are increasingly concentrated around percolating energy contours \cite{Isichenko1992}. For this reason, nontrivial exponents associated with 2D percolation enter the expressions for the resistivity, giving a stronger temperature dependence $\rho \propto T^{10/3}$ \cite{HuiPozderac2024}.

\subsection{Current noise of hydrodynamic electrons}
\label{sec: noise}

Electric current through a material is inevitably accompanied by random temporal fluctuations about its average value. Such current noise can generally be classified into two types: shot noise and Johnson-Nyquist noise. 

Shot noise is associated with the discreteness of the charge carriers. To understand its origin in a slightly oversimplified way, consider a current $I$ flowing through a resistor. During a time interval $\Delta t$, this current is associated with a certain number $N = I \Delta t/q$ of charge carriers moving from the source to the drain electrodes. If the charge carriers move independently of each other at some finite rate, then there are statistical fluctuations in $N$ over any particular time interval, and these are of order $\Delta N \sim \sqrt{N}$; the arrivals are described by a Poisson process. Consequently, there are statistical fluctuations in the average current over that time interval of order $\Delta I = q \Delta N / \Delta t$. The noise power $(\Delta I)^2$ is therefore given by $q^2 N/(\Delta t)$, or $(\Delta I)^2 \sim |q| I \Delta f$, where $\Delta f \sim 1/\Delta t$ is the frequency bandwidth over which the current fluctuations are measured \footnote{The shot noise is sometimes quoted with an extra prefactor of $2$, i.e. $S(\omega) \equiv \langle I(\omega) I(0) \rangle = 2eI$. This depends on the convention of Fourier transform; i.e. if one is computing the full two-sided Fourier transform (as done here) it does not have this factor of $2$. However, if one is considering the one-sided Fourier transform or spectral noise density (as is commonly done in the literature), one obtains an additional factor of $2$. \label{fn: shot noise fourier convention}}. Thus, shot noise as an experimental tool gives a direct measurement of the quasiparticle charge $q$ (see, e.g., \cite{deJong1997, Blanter2000} for general reviews about shot noise). Shot noise measurements have been used, for example, to confirm the existence of fractionally charged quasiparticles in the fractional quantum Hall effect \cite{samindayar1997, dePicciotto1998}. 

However, even when no dc current is flowing, there are still statistical fluctuations about $I = 0$ that are associated with thermal fluctuations in the velocity of charge carriers. The corresponding current noise is called Johnson-Nyquist (JN) noise \cite{johnson_thermal_1928, nyquist_thermal_1928}. The classical form of JN noise is $(\Delta I)^2 = (2 k_B T / R) \Delta f$, which can be thought of as one of the many incarnations of the fluctuation-dissipation theorem \footnote{Again, this is sometimes written with an additional factor of $2$ due to Fourier transform conventions. See footnote~\ref{fn: shot noise fourier convention}.}. JN noise measurements are also a powerful experimental tool, since they provide a direct readout of the electron temperature (provided that one separately measures the two-terminal resistance $R$). Such noise thermometry can be exploited to make sensitive measurements of the electronic thermal conductivity and specific heat, which are accomplished by depositing a known amount of heat into the electron system and measuring its increase in temperature through JN noise \cite{Fong2012, Crossno2015, Crossno2016, Waissman2022}. JN noise measurements also allow one to construct ultrasensitive light detectors (bolometers), in which the temperature increase of electrons that results after light absorption is detected through an increase in JN noise \cite{Karasik2014, Efetov2018, Miao2018, Liu2018, Miao2021}.

In all of these applications, the sensitivity of the measurement is improved when the electrical conductivity is large and the thermal conductivity is small, so that heat is easily absorbed by the electron system and is not easily conducted away to the contacts. In other words, practical uses of JN noise benefit from the electron system having a small Lorenz number $L_\text{WF}$ [see Eq.~\eqref{eq: WF}]. As explained above, hydrodynamic electron systems (with a single species of carriers) in general have a greatly reduced value of $L_\text{WF}$, so that electron systems in the hydrodynamic regime are ideal for applications using JN noise.

A key theoretical challenge for interpreting JN measurements in the hydrodynamic regime has been the question of how to relate local thermal fluctuations in electron velocity to the global current collected between the source and drain electrodes, particularly given that the electron temperature is nonuniform when heated by light absorption or by Joule heating. For ohmic transport, this relation is provided by the so-called Shockley-Ramo theorem \cite{Shockley1938, Ramo1939}, which provides a powerful and general recipe for relating a local current source to a measured global current \cite{Song2014, Pozderac2021}. The Shockley-Ramo approach allows one to draw very general conclusions about the relation between the heat deposited in the electron system and the increase in JN noise \cite{Pozderac2021}, but this generality arises from the fact that both the electrical and thermal currents are dictated by a local conductivity tensor ($\sigma$ and $\kappa$, respectively). In hydrodynamic electron systems, however, the thermal conductivity is local but the electrical conductivity is nonlocal. As a result, the interpretation of experiments on hydrodynamic materials must make reference to the geometry of the sample and involves solving a nontrivial calculation related to the viscous heat dissipation.

The proper generalization of the Shockley-Ramo theorem to the hydrodynamic context was developed in Refs.~\cite{Hui2023, Hui2024}. This generalization involves treating the correlations in electron velocity using the hydrodynamic equations presented in Sec.\ \ref{sec: framework} and properly generalizing the boundary conditions discussed in Sec.~\ref{sec: corbino}. In the case of a rectangular sample, the ratio between the dissipated power and the increase in JN noise is surprisingly similar in all regimes to the naive result associated with the Shockley-Ramo theorem \cite{Hui2023}. This similarity enabled \textit{a posteriori} justification of the experimental interpretation in Ref.~\cite{Crossno2016}, which used JN noise measurements to infer the breakdown of the WF law in the hydrodynamic regime.

\begin{figure}
    \centering
    \includegraphics[width=.9\linewidth]{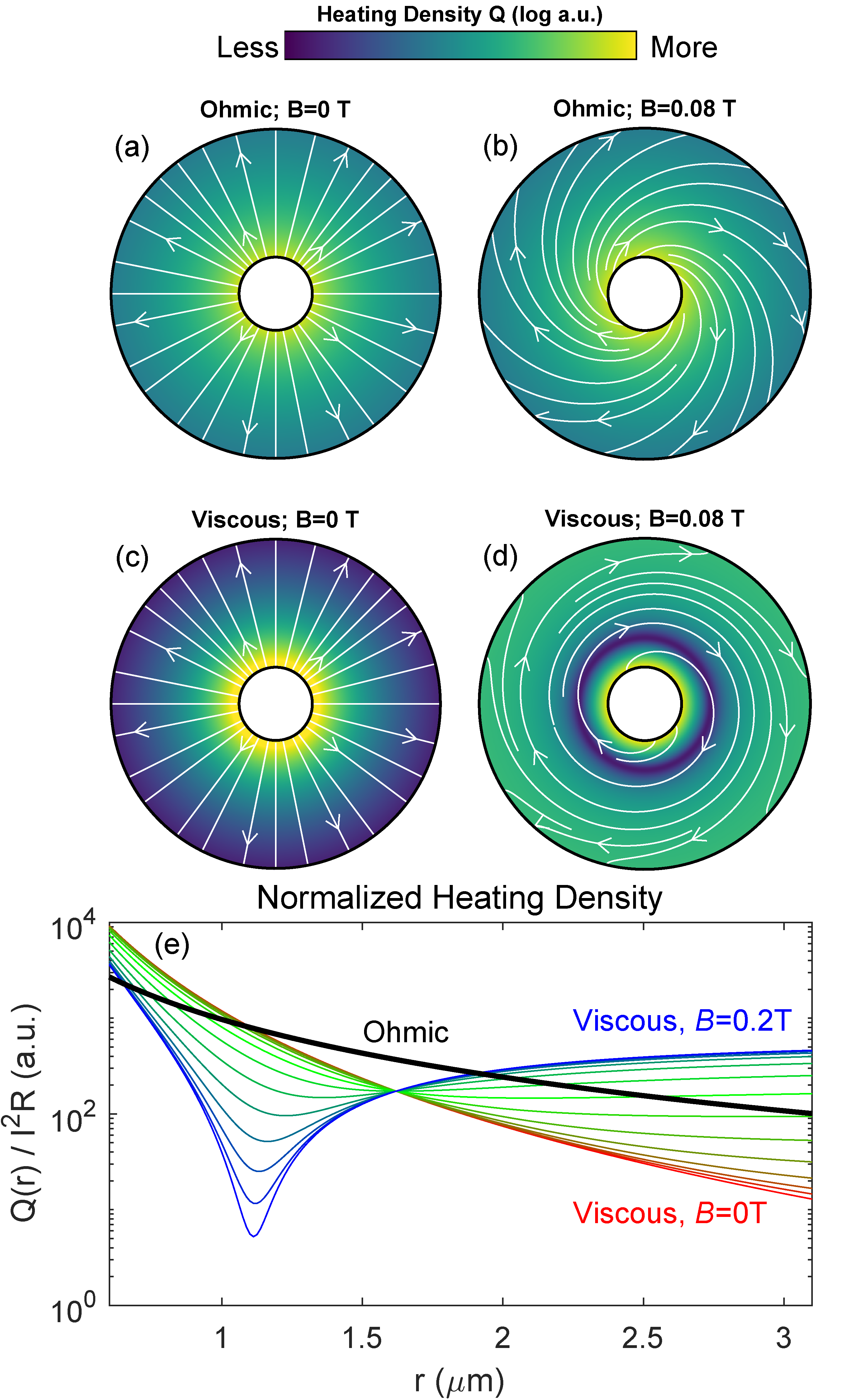}
    \caption{(a-d) Plots of the (normalized) heating profiles in the ohmic and viscous limits, with the white arrows denoting the flow streamlines. In the ohmic case (a-b), there is no effect of magnetic field. In the viscous case (c-d), there is a topographical difference in heating; under magnetic field, the meating is minimal in the center. (e) A radial line-cut of the the normalized heat profiles. As previously mentioned, under magnetic field the heating profile develops a minimum in the center of the device. Adapted from Ref.~\cite{Talanov2024}.}
    \label{fig: Corbino heating and temp}
\end{figure}

The temperature profile of the electron system with an applied current can also be strongly different from the case of ohmic flow, since viscous shear provides a source of heating that is qualitatively dissimilar from Joule heating in ohmic flow \cite{Hui2023, Hui2024}. In Ref.~\cite{Talanov2024}, it was shown experimentally that this difference provides a new qualitative signature of electron hydrodynamics. Namely, one can measure the JN noise amplitude under the application of a perpendicular magnetic field in a Corbino device. In the ohmic regime, the magnetic field increases the electrical resistance, leading to more Joule heating and a higher electron temperature. However, when normalized by input power, the heating profile under magnetic field is identical to the one without magnetic field (see Fig.~\ref{fig: Corbino heating and temp}(a-b)). Things behave quite differently in the hydrodynamic regime. Here, the dominant source of heating arises from viscous shear, and in the presence of magnetic field this shear occurs increasingly close to the inner contact as electric current swirls under the influence of the magnetic field (see Fig.~\ref{fig: Corbino heating and temp}(c-d)). Consequently, the heat deposited into the electron system is more easily dissipated into the contact and the apparent temperature of the electron system, as measured by JN noise, is reduced. This ``apparent negative thermal magnetoresistance'' is a qualitative signature of the hydrodynamic regime. Fitting the experimental data also provides a novel method to estimate the electron viscosity using only two-terminal measurements \cite{Talanov2024}.

The behavior of shot noise in the hydrodynamic regime is less well understood. For hydrodynamic electrons, there is, naively, no notion of charge carriers moving independently of each other as they carry current, so that one can expect statistical fluctuations in the current to be suppressed by electron correlations. Preliminary data suggests that the magnitude of shot noise (quantified by the so-called Fano factor) decreases as one moves closer to the hydrodynamic regime \cite{talanov2023hydrodynamics, Nagaev1995}, but this effect remains incompletely explored. Theoretical works have explored the suppression of shot noise near a quantum critical point, where there are no well-defined electronic quasiparticles \cite{Nikolaenko2023, Wang2024}, following observations of suppressed shot noise in the heavy fermion compound YbRh$_2$Si$_2$ \cite{chen2023shot}. However, to our knowledge there is still no theoretical treatment of shot noise in hydrodynamic electron systems.

\subsection{Nonlinear effects}

Although the full Navier-Stokes-Ohm equation is nonlinear, most work has focused on the linear Stokes-Ohm regime. Nonlinear effects require sufficiently large velocities, as can be measured by the Reynolds number and its related counterparts. The Reynolds number is defined by estimating the ratio of the convective force to the viscous force, namely
\begin{align}
    \operatorname{Re} \equiv \frac{uL}{\nu}
\end{align}
where $u$ is a characteristic velocity, $L$ is a characteristic length, and $\nu$ is the kinematic viscosity. For the Navier-Stokes-Ohm equation, another dimensionless ratio also appears, namely the ``momentum-relaxation'' Reynolds number \cite{Hui2021} defined by
\begin{align}
    \operatorname{Re}_\text{mr} \equiv \frac{u}{\gamma_\text{mr} L}
\end{align}
arising from the ratio of the momentum-relaxation term with the convective force, in analogy to the usual Reynolds number. For nonlinear effects to become important, both $\operatorname{Re}$ and $\operatorname{Re}_\text{mr}$ must be sufficiently large. In classical fluid dynamics, turbulent behavior begins around $\operatorname{Re} \sim 10^3$, though nonlinear effects can appear at much lower $\operatorname{Re} \sim 1$ \footnote{We are being approximate and schematic here regarding Reynolds numbers; the exact transition point to turbulence depends on the specific geometry in question and can vary widely.}. Unfortunately, current experiments can only reach $\operatorname{Re}, \operatorname{Re}_\text{mr} \sim 10^{-2}$ before risking obliteration of the device from thermal heating. This makes probing nonlinear effects challenging.

Low Reynolds numbers notwithstanding, there have been a few theoretical works exploring nonlinear hydrodynamic effects in electron systems. Perhaps the simplest nonlinear effect is the classical Bernoulli effect. In the absence of dissipation, conservation of energy implies that
\begin{align}
    \frac{1}{2}\rho u^2 + e\phi = \text{const}
\end{align}
which is known as the Bernoulli equation. Despite its simplicity, the equation is nonlinear in $u$. For an incompressible flow where the inlet and outlet widths are not identical, e.g.\ a Venturi tube geometry, this nonlinearity gives rise to a non-linear and singular $I$-$V$ characteristic with $I \propto \sqrt{V}$ arising from convective acceleration \cite{Hui2021}. This $I$-$V$ characteristic has a singular point, below which may exist a potential instability towards turbulent and/or intermittent flow. Typical nonlinear perturbative calculations of $I$ in powers of $V$ may have difficulty recovering this result due to the square-root singularity; the hydrodynamic approach, however, displays this nonlinearity straightforwardly.

Nonlinearities also give rise to harmonic generation and mode-mixing effects, as the product of two sinusoids at frequencies $\omega_1$ and $\omega_2$ is equivalent to sum of two sinusoids at frequencies $\omega_1 \pm \omega_2$. Some simple examples of this include Eckart and Rayleigh streaming as well as photon drag, where the convective acceleration term provides DC rectification of an input AC signal \cite{Dyakonov1996, Tomadin2013, Sun2018, Hui2021}. Furthermore, as a quadratic nonlinearity, the convective acceleration gives rise to second harmonic generation. These effects can potentially be used as nonlinear circuit elements for future technological applications, such as for hydrodynamic photovoltaic devices or for detection of THz radiation \cite{Dyakonov1996, Tomadin2013, Hui2021}.

One of the more dramatic harmonic generation effects is the Dyakonov-Shur instability, which has been proposed as a means of THz generation \cite{Dyakonov1993, Mendl2018, Mendl2021, Crabb2021, Farrell2022}. The basic setup requires asymmetric boundary conditions, with one of the two contacts (say, the source) being held at a fixed voltage (relative to a gate electrode) while a current source maintains a fixed bias current into the drain. The fixed-voltage BC causes reflection of incident electronic waves, and fluctuations of the current at the source are possible even for a fixed bias current due to the shunting of current between the source and gate electrodes.
% The current-fixed BC determines a overall bias current in the device, while the voltage-fixed BC reflects any incident wave.
When an incident wave travels from source to drain, the reflected wave is given a different speed than the incident wave as a result of the drift induced by the bias current. If the bias current is sufficiently large, then the reflected wave moves in the same direction as the incident wave; this is the Dyakonov-Shur amplification instability. The Dyakonov-Shur instability has long been of technological interest as a way of addressing the so-called THz gap, which is an underutilized frequency band from $0.1$ to $10$ THz in which it is difficult to generate and detect EM waves. Although THz detection via downconversion from hydrodynamic nonlinearities has been achieved \cite{Tauk2006, Giliberti2015}, a direct detection of THz generation is still lacking.

At sufficiently high Reynolds numbers, one can also induce turbulent behavior. By analogy to the classical fluid, one also expects Kolmogorov scaling with a direct energy cascade in 3D \cite{landauv6}, and an inverse energy cascade in 2D \cite{Boffetta2012}. However, the presence of finite $\tau_\text{mr}$ will cut off and potentially spoil the cascading behavior (see e.g.\ Ref.~\cite{Lucas2018review}). Theoretical works have proposed to observe phenomena such as the Rayleigh-B\'enard instability \cite{Furtmaier2015}, the Kelvin-Helmholtz instability \cite{Coelho2017}, and preturbulent behavior via current or electric potential fluctuations \cite{Mendoza2011, Gabbana2018, DiSante2020}. There has also been a proposal to observe the nonlinear dynamics of hot spot relaxation\cite{Briskot2015}. As previously mentioned, however, reaching the large drive parameters (e.g.\ large Reynolds numbers) necessary for these strongly nonlinear effects seems to be experimentally difficult \cite{Sukhachov2021, Lucas2018review, Hui2021}.

\section{Beyond Isotropic Fermi Liquid Hydrodynamics}
\label{sec: beyond}

\subsection{Symmetry-breaking and the viscosity tensor}

In the previous section, we focused on the prototypical example of an isotropic Fermi liquid with $\epsilon_k = \hbar^2 k^2/(2m)$ dispersion. With the exception of Fermi statistics, the treatment of this fluid is identical to that of a classical Boltzmann gas. In solid materials, however, the quasiparticle properties can vary widely. For instance, many symmetries can be broken as compared to the isotropic, time-reversal invariant classical gas. In the absence of symmetry constraints, viscosity is a rank-4 tensor
\begin{align}
   \sigma_{ij}' = \eta_{ijk\ell} \partial_k u_\ell
   \label{eq: full viscosity}
\end{align}
where $\sigma_{ij}'$ is the viscous stress tensor. In the presence of time-reversal symmetry, one expects $\eta_{ijk \ell} = \eta_{k\ell ij}$ as a consequence of Onsager reciprocity. Rotational invariance further constrains the viscosity tensor to be described by two components of bulk and shear viscosity. Both of these can easily be broken in solid state systems.

Rotational invariance is obviously broken by the crystal lattice; in fact, it is not a microscopic property of any solid state system. Instead, rotational invariance can only be a low-energy emergent symmetry as reflected by the shape of the Fermi surface, such as in graphene near the Dirac point. Many systems are anisotropic, and thus require more non-trivial tensor components to describe viscosity. These components have been studied a number of works \cite{Cook2019, Varnavides2020, Rao2020, Epstein2020, Cook2021, Qi2021, Khain2022, Afanasiev2022} though have yet to be experimentally measured. 

Analogously, time-reversal symmetry is easily broken in these systems by the application of a magnetic field. This gives rise to a Hall viscosity or odd viscosity \cite{Avron1995, Avron1998}, a dissipationless contribution to viscosity much like the Hall resistivity. The Hall viscosity has long been of interest due to its predicted connection to the topological properties in a quantum Hall fluid (at high magnetic fields), in analogy to the quantization of the Hall conductivity \cite{Read2009, Hughes2011, Read2011, Hoyos2012, Bradlyn2012, Cho2014, Sherafati2016, Alekseev2016, Pellegrino2016, Delacretaz2017, Scaffidi2017, Holder2019, Burmistrov2019, Rao2020}. The Hall viscosity coefficient has been measured in a couple of experiments \cite{Gusev2018, Berdyugin2019}.

\subsection{Electron-Hole and Dirac Fluid Hydrodynamics}
\label{sec: dirac}

In electron systems, the quasiparticles need not be of a single species nor have a parabolic dispersion relation. Monolayer graphene, the most popular hydrodynamic material, has a linear $\epsilon_{k,\pm} = \pm \hbar v_F k$ Dirac dispersion. Near charge neutrality, i.e.\ when the chemical potential $\mu \ll k_B T$ is much smaller than the temperature, both conduction band electrons and valence band holes are thermally excited and contribute to transport phenomena; this regime is known as the Dirac fluid. In bilayer graphene, the dispersion is approximately described by a quadratic band-touching point $\epsilon_{k,\pm} = \pm \hbar^2 k^2/(2m)$. Similarly, both electrons and holes contribute to transport near charge neutrality. The hydrodynamics of these materials thus contains two oppositely-charged quasiparticle species, which we call an electron-hole plasma \cite{Fritz2024}. These modifications to the dispersion lead to a few significant modifications to hydrodynamic behavior, which we describe below. 

One of the surprising features of an electron-hole plasma is that it has a finite, local conductivity even at charge neutrality and even in the absence of impurity or phonon scattering. Precisely at charge-neutrality, a finite temperature yields a cloud of electrons and holes. By applying an electric field, the electron and hole clouds move in opposite directions; the total momentum remains zero, but a finite current emerges. The charge current can be relaxed by momentum-conserving electron-hole scattering; unlike in the clean Fermi liquid with only one sign of charge carriers, momentum and charge current are not proportional to each other and thus Peirel's ``infinite conductivity'' argument is evaded. This zero-momentum conductivity, in which electrons and holes carry momentum in opposite directions, is also sometimes called incoherent conductivity, quantum-critical conductivity, or intrinsic conductivity \cite{Lucas2018review, hartnoll2018book, Fritz2024}. At finite but low density, the usual finite-momentum Drude mode couples to the electric field. In a simplistic relaxation-time approximation model where these two modes are independent, the total conductivity is a sum of contributions from the usual Drude conductivity and the zero-momentum intrinsic conductivity. 

An analogous argument can be constructed to understand the thermal conductivity \cite{Fritz2024}. A temperature gradient results in both electron and hole clouds drifting in the same direction; at charge neutrality, a finite momentum current but zero electric current results as a result of the temperature gradient. For both linear (Dirac) and quadratic band-touching points, it has been shown that the heat current has a strong overlap with the momentum current at charge neutrality. This result is immediate for a Dirac dispersion, since the energy current $\mathbf{j}_E \propto v_F \epsilon_k \propto v_F k \propto \mathbf{g}$ is proportional to the momentum current (and the heat current $\mathbf{j}_Q \propto (\epsilon_k - \mu) v_k$). For quadratic band-touching or other electron-hole plasmas this result has also been demonstrated, although it is less straightforward \cite{Zarenia2019, Zarenia2019b, Zarenia2020, Fritz2024}. As a result, momentum-conserving collisions cannot degrade $\kappa$; the thermal conductivity is controlled only by the momentum-relaxation rate $\gamma_\text{mr}$. At charge neutrality and in the clean limit, this leads to a large enhancement of the Wiedemann-Franz ratio $L_\text{WF} \sim \gamma_\text{imp}/\gamma_\text{ee} \rightarrow \infty$ since $\kappa$ diverges (see also \cite{Foster2009, Muller2008, Fritz2008, Muller2008a, Xie2016, Lucas2016a, Tu2023, Tu2023a, Levchenko2025}) in the limit where there is no momentum relaxation while $\sigma$ remains finite. We remark that in making this approximate argument, we have implicitly ignored any viscous contributions to the total conductance. A large thermal conductivity at charge neutrality in graphene has been experimentally observed \cite{Crossno2016,Gallagher2019, Block2021}.

The above phenomena occur as a result of having two oppositely-charged quasiparticle species and are independent of the details of the quasiparticle dispersion. We now focus on the linear Dirac dispersion of monolayer graphene, which has been the subject of many papers. The linear dispersion explicitly breaks Galilean invariance. The quasiparticles are pseudo-relativistic, and therefore relativistic hydrodynamics has been suggested as a starting point for understanding the Dirac fluid \cite{Lucas2018review}. The Coulomb interactions are not Lorentz-invariant (with $v_F$ playing the role of the speed of light), and therefore the Lorentz symmetry is at best approximate. Regardless, one can always work uncontentiously in the Boltzmann framework to derive the continuity equations and constitutive relations \cite{Svintsov2013, Tomadin2013, Narozhny2015, Briskot2015, Narozhny2019}. 

At the level of linear response, the equations of motion for hydrodynamic Dirac electrons derived from the Boltzmann equation are structurally identical to the clean isotropic Fermi liquid \cite{Narozhny2019, Briskot2015, Lucas2018review}; this equivalence is unsurprising since linear transport can only be described by a non-local Ohm's law. The primary difference is in the microscopic nature of momentum-relaxation and viscosity. The momentum-relaxation rate $\gamma_\text{mr}$ has an additional contribution from electron-hole scattering as discussed above; the intrinsic conductivity is expected to take a universal value in graphene \cite{Damle1997, Hartnoll2007} and has recently been experimentally measured \cite{Gallagher2019, Majumadar2025}. The dynamic shear viscosity is predicted to scale as $\eta \sim T^2$ at charge-neutrality. This result can be qualitatively estimated by $\eta \sim n E_F \tau_{ee}$, where $n \sim T^2$ and $E_F \sim T$ from thermal excitations while the interacting scattering rate takes the quantum-critical value $\tau_\text{ee} \sim 1/T$ \cite{Fritz2008, Muller2009, Narozhny2019}. We also note that interpretation of the hydrodynamic mass is subtle. Although Dirac quasiparticles are massless, the hydrodynamic mass corresponds to the energy and momentum stored in the fluid. In other words, the hydrodynamic mass is defined by the response of a fluid packet to an external force. As a result, the hydrodynamic mass $m$ is related to the local energy density as $\epsilon/v_F^2$ or equivalently is equal to the cyclotron mass \cite{Svintsov2013, Narozhny2019}. Though the mass becomes velocity-dependent as a result, this dependence is a subleading effect in linear response.

At nonlinear order, however, the equations of motion are changed relative to the case of a parabolic dispersion. Due to the linear Dirac dispersion, the fluid is no longer Galilean-invariant, and this lack of invariance leads to corrections to the convective term in addition to other nonlinearities \cite{Svintsov2013, Tomadin2013, Briskot2015, Narozhny2019, Sun2018, Cosme2023, Lucas2018review}. Predictions regarding modifications to plasma waves and oscillations have been made, though detailed experimental confirmation is still lacking. We remark that a recent experiment \cite{Zhao2023} has observed the relativistic sound mode (demon mode or energy wave) with characteristic speed $v_F/\sqrt{2}$ in graphene \cite{Lucas2018review, Briskot2015, Sun2018, Fateev2020}, though this effect arises from linear response. This mode is analogous to second sound in typical Fermi liquids (e.g. liquid He-3), which is also an energy wave of speed $v_F/\sqrt{d}$ for space dimension $d$.

We briefly remark that there are a number of works that take a slightly different approach to constructing the equations of motion near charge neutrality \cite{Li2020, Li2022, Li2022a, Levchenko2022, Levchenko2022a}. These works treat the decomposition of zero-momentum and finite-momentum modes honestly, phenomenologically constructing a standard ohmic transport description for the zero-momentum mode while writing a Stokes equation for the finite-momentum ``hydrodynamic'' mode. This approach results in slightly different equations of motion as compared to constructing hydrodynamics from the Boltzmann equation and then taking the appropriate limits. At the time of writing, it is unclear to the authors how similar or distinct these two approaches are.

\subsection{Electron-Phonon Hydrodynamics}

So far we have described electron-phonon scattering merely as a source of momentum relaxation, such that phonons take momentum from the electron system and dissipate it elsewhere. As such, we have considered such scattering as detrimental to the formation of a hydrodynamic electron fluid. However, if the phonons relax momentum slowly, they can also transfer momentum back into the electron system. As such, the combined electron-phonon system could be momentum-conserving and be in a hydrodynamic regime, even though the electron component alone may not be hydrodynamic. Analogous to the electron-hole fluid, the electron-phonon fluid is comprised of two quasiparticles, electrons and phonons. Just as before, one can use the same Boltzmann approach to derive continuity equations and constitutive relations. As the mathematical formalism is much the same, we only briefly comment on this regime and leave detailed description of the results to Refs.~\onlinecite{Levchenko2020, Huang2021, Fritz2024}. 

The most dramatic qualitative effect of electron-phonon hydrodynamics is phonon drag -- if the electron-phonon fluid is strongly coupled, the electrons and phonons will tend to move together. Thus, upon applying an electric field, the charged electrons experience a driving force and drag the neutral phonons along. On the other hand, a temperature gradient produces a driving force for both electrons and phonons. This asymmetry leads to violations of the WF law, as well as an enhancement of thermoelectric effects. (Hydrodynamic phonons interacting with electrons have also been long suggested to produce interesting forms of second sound \cite{gurevich1966, nielsen1968}.) However, as previously mentioned for the clean Fermi liquid, observations of these effects must be done with care to exclude other possible pathways to the same effect. Experimentally, electron-phonon hydrodynamics has been proposed for Sb \cite{Jaoui2022}, WTe$_2$ \cite{Vool2021}, WP$_2$ \cite{Jaoui2018, Gooth2018}, ZrTe$_5$ \cite{Galeski2024}, PtSn$_4$ \cite{Fu2018, Fu2020}, and NbGe$_2$ \cite{Yang2021}, though detailed confirmations between theory and experiment have yet to be done. 

\section{Conclusion and Outlook}
\label{sec: conclusion}

The observation of the hydrodynamic regime and the treatment of electron-electron interactions have led to a reevaluation of traditional transport lore. The transport properties of a clean, hydrodynamic Fermi liquid are qualitatively different in many regards from the usual momentum-relaxing metallic Fermi liquid. For instance, even the basic concept of a local resistivity fails in the hydrodynamic material and the ``fundamental'' Landauer bound on conductance is less fundamental than originally thought. Such phenomena have motivated not only an interest in novel materials, but also a resurgence of interest in long-studied 2DEGs with a more careful eye towards hydrodynamic effects \cite{deJong1995, Gusev2018, Levin2018, Braem2018, Keser2021, Gupta2021, Gusev2021, Vijayakrishnan2023, Palm2024}. Though many of the effects described have a classical fluids analogue, we have noted some non-trivial extensions of hydrodynamic behavior in electronic systems in Sec.~\ref{sec: beyond}. Furthermore, one can also extend the hydrodynamic or kinetic theory approach to include quantum or topological effects, such as Berry curvature \cite{Hasdeo2021, Grigoryan2024}. 

Throughout this review, we have characterized the hydrodynamic regime as ``strongly-interacting.'' However, we have mostly contented ourselves with the regime where interactions are strong relative to momentum-relaxing scattering processes but still sufficiently weak that a Boltzmann description still applies, i.e.\ the interactions can still be treated perturbatively. To draw an analogy to classical fluids, these are ``gas-like'' systems. However, there are also many ``liquid-like'' systems where interactions are non-perturbatively strong, e.g.\ systems where quasiparticles are short-lived and the Boltzmann approach fails. Even in the classical case, the behavior of liquids can be very different from those of gases. For instance, viscosity tends to decrease with temperature in a classical liquid while viscosity increases for a classical gas. This observation has led to classical conjectures on the possibility of a lower bound on viscosity \cite{Purcell1977, Trachenko2020} and a quantum counterpart originating from the AdS/CFT literature \cite{Kovton2005}. In general, strongly-correlated (``liquid-like'') systems are difficult and nigh-intractable to study from a first-principles approach.

The surprising power of the hydrodynamic framework is that it works equally well for classical liquids as it does for classical gases, despite the inability to derive the equations, e.g. from a Boltzmann framework. One could also hope that the hydrodynamic framework may provide a new window into long-standing transport mysteries. The strongly correlated strange metal phase is one such example, which has a low resistivity $\rho \sim T$ that mysteriously scales linearly with $T$ \cite{Bruin2013}. Several works have suggested links to hydrodynamic behavior \cite{Davidson2014, Hartnoll2015, Lucas2015, Lucas2017, HuiPozderac2024}, though its true origin is still unclear. The power of the hydrodynamic framework is to unify, in a non-perturbative way, the language with which we discuss transport behavior in a wide array of systems. In so doing, we not only uncover novel transport phenomena but hope that this framework may shed some light on long-standing mysteries in strongly correlated transport.

\section{Acknowledgments}
We thank Patrick Ledwith, Sankar Das Sarma, and Boris Shklovskii for helpful comments regarding this review. We thank Yihang Zeng, Cory Dean, and Felicia Setiono for graciously providing figures. B.S. was supported by the NSF under Grant No. DMR-2045742.

\bibliography{biblio}

\end{document}